\documentclass[12pt]{article}
\usepackage{amsmath, amssymb}
\usepackage{graphicx}
\usepackage{bm}
\RequirePackage{tikz}
\usepackage{enumerate}
\usepackage[colorlinks,citecolor=blue]{hyperref}
\usepackage{color}
\usepackage{appendix}
\usepackage{mathrsfs}
\usepackage{float}
\usepackage{algorithm}
\usepackage{algpseudocode}
\usepackage{bigstrut}
\usepackage{ulem}
\usepackage{dsfont}
\usepackage{cleveref}
\usepackage{subcaption}
\usepackage{booktabs}
\usepackage{pdflscape}
\usepackage{multirow}
\usepackage{natbib}
\usepackage{ntheorem}
\usepackage{authblk}
\usepackage{xr}
\usepackage{url} 
\usepackage{enumitem}
\usepackage[top=1in,bottom=1.3in,left=1in,right=1in]{geometry}

\graphicspath{{figures/}}

\newcommand{\E}{\mathbb{E}}

\newcommand{\tdomain}{\mathcal{X}}

\def\diff{{\rm d}}
\def\argmin#1{\mathrel{\mathop{\arg\min}\limits_{#1}}}

\newcommand{\bea}{\begin{eqnarray*}}
\newcommand{\eea}{\end{eqnarray*}}
\newcommand{\be}{\begin{eqnarray}}

\newcommand{\ee}{\end{eqnarray}}
\newcommand{\bsp}{\begin{split}}
\newcommand{\esp}{\end{split}}
\newcommand{\ed}{\end{document}}

\newcommand{\btab}{\begin{tabular}}

\newcommand{\etab}{\end{tabular}}
\newcommand{\bc}{\begin{center}}
\newcommand{\ec}{\end{center}}

\newcommand{\bi}{\begin{itemize}}
\newcommand{\ei}{\end{itemize}}
\newcommand{\bfi}{\begin{figure}}
\newcommand{\efi}{\end{figure}}
\newcommand{\ben}{\begin{enumerate}}
\newcommand{\een}{\end{enumerate}}
\newcommand{\bdes}{\begin{description}}
\newcommand{\edes}{\end{description}}

\newcommand{\bay}{\begin{array}}
\newcommand{\eay}{\end{array}}

\newcommand{\RN}[1]{%
  \textup{\uppercase\expandafter{\romannumeral#1}}%
}

\def\bco{\iffalse}

\def\var{{\rm var}}

\def\diag{{\rm diag}}

\def\cp{\citep}

\def\F{Fr\'echet }

\def\M{\mathcal{M}}

\newcommand\independent{\perp\!\!\!\perp}
\newcommand\nindependent{\not\!\perp\!\!\!\perp}

\def\bco{\iffalse}

\def\var{{\rm var}}

\def\reall{{\mathbb{R}}}

\def\diag{{\rm diag}}

\def\cp{\citep}

\def\F{Fr\'echet }

\newcommand{\blind}{1}

\theoremstyle{plain}
\theoremheaderfont{\scshape}
\theoremseparator{.}
\theorembodyfont{\itshape}
\newtheorem{theorem}{\indent\large Theorem}
\newtheorem{prop}{\indent\large Proposition}
\newtheorem{corollary}{\indent\large Corollary}

\newtheorem{assumption}{\indent\large Assumption}

\makeatletter
\newtheoremstyle{proof}%
{\item[\hskip\labelsep\theorem@headerfont\MakeUppercase ##1\theorem@separator]}%
{\item[\hskip \labelsep\theorem@headerfont\MakeUppercase ##1\ ##3\theorem@separator]}
\makeatother

\theoremstyle{proof}
\theoremheaderfont{\scshape}
\theorembodyfont{\upshape}
\theoremsymbol{\ensuremath{\square}}

\theoremseparator{}

\begin{document}

	\def\spacingset#1{\renewcommand{\baselinestretch}%
		{#1}\small\normalsize} \spacingset{1}

	
	\if1\blind
	{
		\title{\bf Association and Independence Test for Random Objects}
		\author[1]{Hang Zhou}
		\author[2]{Hans-Georg M\"{u}ller}
		\affil[1]{Department of Statistics and Operations Research \& School of Data Science and Society, University
of North Carolina at Chapel Hill}
		\affil[2]{Department of Statistics, University of California, Davis}
		\maketitle
	} \fi

	\bigskip
	\begin{abstract}
		We develop a unified framework for testing independence and quantifying association between random objects that are located in general metric spaces. Special cases include functional and high-dimensional data as well as networks, covariance matrices and data on Riemannian manifolds, among other metric space-valued data.
		A key concept is the profile association, a measure based on distance profiles that intrinsically characterize the distributions of random objects in metric spaces. 
		We rigorously establish a connection between the Hoeffding D statistic and the profile association and derive a permutation test with theoretical guarantees for consistency and power under alternatives to the null hypothesis of independence/no association.  
		We extend this framework to the conditional setting, where the independence between random objects given a Euclidean predictor is of interest.
		In simulations across various metric spaces, the proposed profile independence test is found to outperform existing approaches.
		The practical utility of this framework is demonstrated with applications to brain connectivity networks derived from magnetic resonance imaging and age-at-death distributions for males and females obtained from human mortality data.

	\end{abstract}
	
	\bigskip
	\noindent%
	\vfill
	
	\newpage
	\spacingset{1} 
	
\section{Introduction}\label{s:intr}

Quantifying association and testing the independence between two random variables is a classical problem at the very core of statistics.  The importance of this problem across many domains, including finance \citep{broock1996test, fan2013large},  genetics \citep{pritchard2000inference, storey2003statistical, browning2007rapid} and many other areas is well known,  and understanding  the intricate relationships between variables is crucial for inference, prediction, and decision-making.
Classical measures of dependence, such as Pearson's correlation \citep{pearson1900mathematical}, Spearman's $\rho$ \citep{spearman1904proof} and Kendall's $\tau$ \citep{kendall1938new} have long served as fundamental tools for analyzing relationships between variables and are still very popular as they are simple and effectively capture linear or monotonic associations.
Recently, the Chatterjee rank correlation coefficient \citep{chatterjee2021new} has garnered significant attention and has been studied and applied across various fields \citep{bick:22, lin2023boosting, han2024azadkia, han2024faliure}.

As data collected in scientific research have become increasingly complex, reaching beyond traditional scalar or vector formats, there is a need for statistical methodology and especially inference that can be applied for data located in more general spaces \citep{dryden2015geometry, dubey2020functional, marron2021object, dubey2022depth}.
Notable examples for such non-traditional and non-Euclidean data are distribution-valued data and Riemannian manifold-valued data, which have numerous applications in fields such as economics \cp{gunsilius2023distributional},  biology \cp{bolstad2003comparison},  brain imaging \cp{lila2016smooth, lila2020statistical} and geography \citep{mitchell2025object}. 
For data residing in  spaces that are endowed with a (pseudo-)Riemannian geometry \citep{o1983semi}, a common approach is to apply transformations, such as the Riemannian  log map, to project the data into a linear tangent space \citep{petersen2016functional, bigo:17, lin2019intrinsic, chen2021wasserstein}.
If such a linearizing map is available, one may adopt traditional methodologies for data in Euclidean spaces such as  quantifications for association of  multivariate or functional data with their usual
Euclidean (or $\mathcal{L}^2$)  metrics.
However, the linearizing maps that are commonly employed are either metric-distorting or non-bijective, requiring ad hoc solutions to construct pseudo-inverses  \citep{bigo:17, chen2021wasserstein}. Also data may reside in metric spaces without an underlying Riemannian geometry, such as the space of phylogenetic trees \citep{billera2001geometry}.

The study of data from general metric spaces has given rise to a prominent research area that has been referred to as metric statistics \citep{dubey2022depth}, where random elements taking values in general metric spaces have been referred to as  random objects \citep{mull:16:7,bhat:25}.
The concept of a mean in general metric spaces was introduced by \cite{frechet1948elements} and more recently extended to define conditional means in regression settings \citep{petersen2019frechet} under the notion of \F regression. 
These concepts are based on intrinsic metric functions and do not require any transformation to an extrinsic linear space.  Such intrinsic approaches that directly confront the absence of linear operations have gained much interest in recent studies \citep{dubey2020functional, zhu2021autoregressive,scho:25,jiang2024two}.
Distance profiles  \citep{dubey2022depth,wang:23:3} corresponding  to a class of one-dimensional distribution functions that characterize the distribution of random objects under mild conditions  have  proved instrumental for   predictive conformal  inference  \citep{zhou2024conformal}.

In this paper, we consider the setting where $X$ and $Y$ are random objects that may be situated in  different metric spaces and develop a unified framework to quantify  the association between $X$ and $Y$.
We extend the concept of distance profiles to random object pairs $(X,Y)$ and demonstrate that the joint distance profile function equals the product of the marginal distance profiles of $X$ and $Y$ if and only if $X$ and $Y$ are independent.
Building on this basic result,  we introduce the \emph{profile association}, a dependence measure that  is zero if and only if $X$ and $Y$ are independent and is one when $X$ and $Y$ have a specific functional relationship that we characterize below.
The analysis of the empirical profile  association relies on  a connection between the Hoeffding D statistic and the proposed profile association, which makes  it possible to devise  a consistent permutation test  for testing the independence between $X$ and $Y$ and to derive its  power properties, the {\em profile independence test}.   Popular alternatives to obtain inference with  distance-based tests include  the energy test and ball divergence test, which  have been extended to non-Euclidean data \citep{szekely2007measuring,lyons2013distance, sejdinovic2013equivalence, wang:23:3, gao2025studentized} and we include these methods in our numerical comparisons.

{
Testing independence between non-Euclidean data has become an active area of research with important applications across many interdisciplinary fields. 
For random elements taking values in Banach spaces, \cite{pan2020ball} introduced an independence measure based on the concept of ball covariance, which is closely related to distance profiles. 
However, their proof of the independence-zero equivalence relies on the linear structure of Banach spaces. In practice, it is uncommon to encounter data that reside exclusively in Banach spaces without also belonging to a Hilbert space, where one can leverage inner product structures and apply a wide range of linear methods \citep{lee2020testing, lundborg2022conditional}.
Recently, \cite{wang:23:3} extended the notion of ball covariance to general metric spaces and included a subsection on its application to independence testing, along with an analysis of the asymptotic distribution of the test statistic under the null hypothesis. 
However, the overall framework for independence testing with random objects remains incomplete: key properties such as the consistency of the test and the power analysis of the permutation-based implementation have not yet been fully established.
}

{
We systematically study the independence testing problem for random objects based on distance profiles.
Intrinsically, both the metric distribution functions (MDF) \cite{wang:23:3} and the distance profiles \cite{dubey2022depth} rely on distribution functions that characterize probabilistic information about distances in a metric space.
In the context of independence testing, there are at least three advantages of using distance profiles over MDF.
First, the framework of distance profiles naturally connects the independence of random objects with the independence of a class of one-dimensional random variables and satisfies a factorization property (see Proposition \ref{c:indP} in Section \ref{s:asso}). This connection bridges the gap between dependence structures of complex objects and those of univariate random variables, allowing one to apply a wide range of existing tests for univariate independence.
Second, because distance profiles are univariate distribution functions, it is straightforward to relate them to Hoeffding's $D$ statistic \citep{hoeffding1948non}, enabling the construction of interpretable and computationally efficient test statistics.
Third, the assumptions required for MDF and distance profiles to fully characterize probability measures on metric spaces are different. In \cite{wang:23:3}, it relies on the directionally $(\epsilon, \eta, L)$-limited condition \citep{federer2014geometric}, which is somewhat abstract and often difficult to verify, particularly in infinite-dimensional spaces such as Wasserstein spaces. 
In contrast, under the distance profile framework, this one-to-one correspondence holds under mild assumptions on either the metric space or the probability measure, making the approach more flexible and broadly applicable.
}

{
Very recently and independently of our work, another study on testing mutual independence between random objects was posted on arXiv \citep{chen2025testingmutualindependencemetric}. While this work also utilizes distance profiles, its proposed test statistic is based on estimated distance profiles. 
Our testing framework, developed simultaneously but independently, differs in several key aspects.
It is based on a Hoeffding's $D$-statistic-type test, establishes the dependence-1 equivalence, employs different proof techniques and employs a half permutation test to study consistency and power.
Furthermore, we extend our approach to the important problem of testing and quantifying conditional associations, conditioning on a Euclidean random vector.
}

The quantification of conditional associations and inference  is motivated by the fact that many real-world datasets involve complex random objects along with Euclidean predictors, such as age, time or experimental conditions.
Understanding conditional relationships  enables researchers to model, predict, and interpret such data effectively, especially in  fields such as  medical imaging, neuroscience, genetics or environmental science.  Specifically, we introduce the notion of  \emph{conditional profile association}  to quantify the dependence between random objects $X$ and $Y$ given a Euclidean predictor $Z$. 
Using estimated joint and marginal distance profile functions, we also propose a  statistic to carry out a  
{\em  conditional profile  independence test} to infer 
 whether $X$ and $Y$ are conditionally independent  and establish the  uniform convergence of the estimated joint and marginal conditional distance profiles, which then implies  the uniform convergence of conditional profile association. We  also study the  asymptotic distribution of the test statistic for the test for conditional independence  under the null hypothesis and its implementation  with a permutation test. 
We evaluate the finite-sample performance of the proposed conditional test through simulation studies and an application to human mortality data, where we examine conditional profile associations between male and female age-at-death distributions across different calendar years (see Section \ref{s:hls}).

In summary, in  this paper we introduce \begin{itemize} \item A  unified framework for testing and quantifying association between random objects with theoretical guarantees for the proposed profile association and profile independence test. 

\item A  half-permutation test procedure that is consistent under both the null and alternative hypotheses. We analyze its asymptotic power and demonstrate that it achieves a root-n minimax separation rate.

\item  An extension to quantify and infer conditional association of random objects in the presence of Euclidean (scalar or vector) covariates. We establish  uniform convergence rates for estimators for the conditional profile  association, derive the asymptotic distribution of the  conditional profile independence test statistic under the null hypothesis and study its implementation with a permutation test. 

\item A demonstration  of the practical merits of the proposed profile independence test and profile association measure through comprehensive simulation studies, demonstrating improved power performance compared to  other  existing methods, complemented by applications to brain connectivity network data and human mortality data. We  establish the  practical utility of the proposed conditional association measure and demonstrate that the proposed approach  is not sensitive to the choice of the underlying metric.

\end{itemize}

The paper is organized as follows. In Section \ref{s:asso}, we introduce the profile  association for random objects along with some key properties. Section \ref{s:meth} presents the main methodology  and theoretical results for empirical versions.  We also discuss  the proposed half permutation test for random objects and its asymptotic properties. In Section \ref{s:cond}, we introduce test statistics and present asymptotic results for quantifying conditional independence in the presence of Euclidean covariates.  In Section \ref{s:simu} we report the performance and comparison of the proposed methods for simulated data, followed by applications to brain imaging and human longevity  that  are presented in Section \ref{s:data}. Proofs and additional results are included in the Supplement.

\section{Profile  association for random objects}\label{s:asso}

We state here basic  assumptions and notations. In the following, \(X\) and \(Y\) are random elements that take values in totally bounded separable metric spaces \(\mathcal{M}_{1}\) and \(\mathcal{M}_{2}\) with metrics \(d_{1}(\cdot, \cdot)\) and \(d_{2}(\cdot, \cdot)\), respectively. 
We refer to such random elements as \emph{random objects}. 
We  denote by  \((S, \mathcal{S}, \mathbb{P})\) an underlying  probability space, where \(\mathcal{S}\) is the Borel \(\sigma\)-algebra on \(S\) and \(\mathbb{P}\) is a probability measure, and  by \(\mu_{XY}\) the joint measure of \(X\) and \(Y\), where  \(\mu_{XY}(A \times B) = \mathbb{P}(s \in S : X(s) \in A \text{ and } Y(s) \in B)\) for any Borel measurable sets \(A \subset \mathcal{M}_{1}\) and \(B \subset \mathcal{M}_{2}\).
	The marginal distributions of \(X\) and \(Y\), denoted by \(\mu_{X}\) and \(\mu_{Y}\), are assumed to have finite first-order moments, i.e., \(\int d_{1}(\omega_{1}, x) \, \mathrm{d}\mu_{X}(x) < \infty\) and \(\int d_{2}(\omega_{2}, x) \, \mathrm{d}\mu_{Y}(x) < \infty\) for some (and thus any) \(\omega_{1} \in \mathcal{M}_{1}\) and \(\omega_{2} \in \mathcal{M}_{2}\).
{
We aim to quantify the association between $X$ and $Y$ and to determine  whether \(X\) and \(Y\) are independent, which is equivalent to \(\mu_{XY} = \mu_{X} \times \mu_{Y}\). 
In general metric spaces, linear operations like addition, subtraction and inner product that are fundamental to conventional statistical inference for Euclidean data are not available. Furthermore, in some metric spaces like  $\mathcal{L}^2{[0,1]}$, the absence of a well-defined Lebesgue measure renders density-based inference methods infeasible. 
Since the only structure available in general metric spaces is the distance between two elements, developing intrinsic statistical methodologies that align with the underlying geometry requires working directly with distances.}

{We further note that 
to develop statistical methods for general metric spaces, one needs to restrict attention to a subclass of probability measures that are fully determined by their values on closed balls. Specifically, this means that the values $\mu(\overline{\mathcal{B}}(\omega, r))$ for all points $\omega \in \mathcal{M}$ and all radii $r > 0$  uniquely determines the measure  $\mu$ on all Borel sets, where $\overline{\mathcal{B}}(\omega, r) := \left\{\omega' \in \mathcal{M} : d(\omega, \omega') \leq r \right\}$ denotes the closed ball centered at $\omega$ with radius $r$.
This property is commonly referred to as the \emph{one-to-one correspondence} and was first studied by \cite{christensen1970some}, \cite{davi:71}, and \cite{hoff:75}. 
More recently, it has been investigated by \citet{blanche2016recovering}, with related applications in statistics and probability discussed in \citet{lyons2013distance}, \citet{dubey2022depth}, and \citet{chen2025testingmutualindependencemetric}.}

Distance profiles  \citep{dubey2022depth} provide a natural framework for studying the one-to-one correspondence and for developing inference tools for random objects by establishing an intrinsic connection between such objects and a class of one-dimensional distribution functions.
This general framework has recently  been applied for  the construction of two-sample tests \citep{dubey2022depth, wang:23:3} and for conformal prediction \citep{zhou2024conformal}.
Specifically, for each $\omega_1 \in \mathcal{M}_1$, the \emph{marginal distance profile} with respect to $\mu_X$ is defined as
\begin{equation}\label{d:marD} F^X_{\omega_1}(u) = \mathbb{P}\left(d_1(X, \omega_1) \leq u \right), \, \, \, u \geq 0,
\end{equation}
and the marginal distance profiles $F^Y_{\omega_2}(u)$ for each $\omega_2 \in \mathcal{M}_2$ with respect to $\mu_Y$ are defined analogously.
To quantify the dependence and characterize independence between $X$ and $Y$, we extend the marginal distance profiles in \eqref{d:marD} to a two-dimensional version  by introducing  \emph{joint distance profiles} for each pair $(\omega_1, \omega_2) \in \mathcal{M}_1 \times \mathcal{M}_2$ with respect to $\mu_{XY}$:
\begin{equation}\label{d:jotD} F_{\omega_1,\omega_2}^{XY}(u,v) = \mathbb{P}\left(d_1(X, \omega_1) \leq u,\, d_2(Y, \omega_2) \leq v \right), \quad \, u,v \geq 0. 
\end{equation}
The concept of joint distance profiles is a natural extension of distance profiles and has also been proposed independently in \cite{chen2025testingmutualindependencemetric}.
The following factorization proposition is a fundamental result that enables us to transform the independence testing problem for random objects into testing a class of two-dimensional vectors.

\begin{prop}\label{c:indP}
Suppose that either of the following assumptions holds:
	\begin{itemize}
		\item[(M0).] $\mathcal{M}_{1}$ and $\mathcal{M}_{2}$ are of strong negative type.
		\item[(P0).] The measures $\mu_X$, $\mu_Y$, and $\mu_{XY}$ are assumed to satisfy the doubling condition:
\begin{equation}\label{eq:doubling} \mu\left\{\overline{\mathcal{B}} (\omega,2r) \right\} \leq L_{\mu}\mu\left\{ \overline{\mathcal{B}} (\omega,r)\right\} \text{ for all }r>0\text{ and } \omega\in\mathrm{supp}(\mu), \end{equation}
for $\mu = \mu_X, \mu_Y,$ and $\mu_{XY}$, where  $\mathrm{supp}(\mu) := \left\{\omega' \in \mathcal{M} : \mu(\{\omega'\}) > 0 \right\}$ denotes the support of $\mu$ and $L_\mu$ is a positive constant. 
	\end{itemize}
Then the random objects  $X$ and $Y$ are independent if and only if  $$
F_{\omega_{1}, \omega_{2}}^{XY}(t, s) = F_{\omega_{1}}^{X}(t) \cdot F_{\omega_{2}}^{Y}(s), \quad \text{for all } t, s \geq 0,\ \omega_{1} \in \mathcal{M}_{1},\ \omega_{2} \in \mathcal{M}_{2}.
$$
\end{prop}
Assumption M0 relates to the structural properties of the underlying metric spaces.
Under this assumption, Corollary 3.20 of \cite{lyons2013distance} shows that the product space $\mathcal{M}_{1} \times \mathcal{M}_{2}$ is also of strong negative type. 
A brief review of the strong negative type assumption  is provided in Supplement, with additional details and  examples available in \citet{kleb:05} and \citet{lyons2013distance}.
Assumption P0 imposes conditions on the probability measures and was adopted in recent work by \citet{chen2025testingmutualindependencemetric} to study mutual independence in metric spaces.
The concept of a doubling measure in metric spaces was first introduced by \citet{beurling1956boundary} and has since become a fundamental notion for metric measure spaces, harmonic analysis and geometric measure theory \citep{heinonen2001lectures}.
The significance of Proposition \ref{c:indP}  is that it reduces the complex independence testing problem for random objects to the simpler problem of studying the dependence of two-dimensional vectors $\{(d_{1}(X,\omega_1), d_{2}(Y,\omega_2))^{\top}\}$, indexed by $\omega_1 \in \mathcal{M}_1$ and $\omega_2 \in \mathcal{M}_2$, where  one can draw on extensive existing research.

{A standard approach for  testing the independence of vector-valued random variables \(X\) and \(Y\) is  to quantify  the discrepancy between their joint measure and the product of their marginal measures. 
For scalar random variables \(X\) and \(Y\) with joint distribution function \(F(x,y)\), in a seminal paper  \cite{hoeffding1948non} introduced the functional 
	\[
	\Delta(F) = \iint \{F(x,y) - F(x,\infty)F(\infty,y)\}^2 \, \mathrm{d}F(x,y),
	\]
	which quantifies the   association  between \(X\) and \(Y\). When \(F(x,y)\) has a continuous density function, \(X\) and \(Y\) are independent if and only if \(\Delta(F) = 0\).}
	
	{
 	When \(X\) and \(Y\) are random objects, one may analogously define their \emph{profile  association} as
\begin{equation}\label{d:indA}
 		\Delta(X,Y):=\E_{X',Y'}\left[ \iint \{F_{X',Y'}^{XY}(u,v)-F_{X'}^{X}(u)F_{Y'}^{Y}(v) \}^2\diff F^{XY}_{X',Y'}(u,v) \right],
 \end{equation}
where \( (X',Y')\) is an independent copy of \((X,Y)\), and the random distance profiles \(F_{X',Y'}^{XY}(u,v)\), \(F_{X'}^{X}(u)\), and \(F_{Y'}^{Y}(v)\) are defined as
\[
\begin{aligned}
    &F_{X',Y'}^{XY}(u,v) := \mathbb{P}_{X,Y}\left(d_1(X, X') \leq u, \, d_2(Y, Y') \leq v \right), \\
    &F_{X'}^{X}(u) := \mathbb{P}_{X}\left(d_1(X, X') \leq u \right), \quad F_{Y'}^{Y}(v) := \mathbb{P}_{Y}\left(d_2(Y, Y') \leq v \right),
\end{aligned}
\]
for all \(u, v \geq 0\). }
Throughout this paper, the subscript in \(\mathbb{E}_{X}\)  (or \(\mathbb{P}_{X}\)) indicates that the expectation (or probability) is taken over the random variable(s) specified in the subscript only. 
For a random object $X$, i.e., a metric space-valued random variable, we denote by $X(s) \in \M_1$ the value it assumes in the metric space $\M_1$ for a specific argument $s$, which is an element of the underlying probability space $S$.
\begin{theorem}\label{thm:assD}
Assume that the density function \( f^{XY}_{\omega_1,\omega_2}(u,v) \) corresponding to the joint distribution function \( F^{XY}_{\omega_1,\omega_2}(u,v) \) exists and is continuous with respect to \( \omega_1 \), \( \omega_2 \), \( u \), and \( v \). Then
\begin{itemize}
	\item[a).] \( \Delta(X,Y) = 0 \) if and only if the random variables \( X \) and \( Y \) are independent.  
	\item[b).]    \( \Delta(X,Y) \) attains its maximum value if and only if the real-valued random variables \( d_{1}(\omega_1,X) \) and \( d_{2}(\omega_2,Y) \) are comonotonic for all \(\omega_1 \in \mathcal{M}_1\) and \(\omega_2 \in \mathcal{M}_2\), i.e., 
\[
\{d_{1}(\omega_1,X(s_1)) - d_{1}(\omega_1,X(s_2))\}\{d_{2}(\omega_2,Y(s_1)) - d_{2}(\omega_2,Y(s_2))\} \geq 0, \text{ for all }  s_1, s_2 \in S.
\]
Furthermore, if \(\mathcal{M}_1 = \mathcal{M}_2\), $X$ and $Y$ are comonotonic if and only if  \(X = Y\) almost surely.  
\end{itemize} 
\end{theorem}
The assumption in Theorem \ref{thm:assD} is relatively mild. Specifically, when \( \mathcal{M}_{1} = \mathcal{M}_{2} = \mathbb{R} \), this assumption can be derived from the condition in \cite{hoeffding1948non}, which requires that the pair \( (X, Y) \) has a continuous joint density function. The first statement of Theorem \ref{thm:assD} shows that the proposed profile association measure satisfies both I-consistency and D-consistency \citep{weihs2018symmetric}:
\begin{itemize} 
\item[\emph{I-consistency}.] If $X$ and $Y$ are independent, the profile   association measure is zero.
 \item[\emph{D-consistency}.] If $X$ and $Y$ are dependent, the profile  association measure is nonzero. 
\end{itemize}
A similar independence-zero equivalence property has been established for ball covariance \citep{pan2020ball}, which was proposed to test independence for the special case of Banach (and thus vector-)space-valued random elements. 
The proof of Theorem \ref{thm:assD} utlizes a well-known mapping between negative-type metric spaces and Hilbert spaces \citep{schoenberg1937certain,schoenberg1938metric, lyons2013distance}.

Theorem \ref{thm:assD} states that \( \Delta(X,Y) \) achieves its maximum value if and only if \( X = Y \) almost surely when $\mathcal{M}_{1}=\mathcal{M}_{2}$. Similar results have been established for other measures of dependence, including  Hoeffding's \( D \) \citep{hoeffding1948non}, Bergsma--Dassios--Yanagimoto's \( \tau^\ast \) \citep{yanagimoto1970measures, bergsma2014consistent, weihs2018symmetric}, and Chatterjee's rank correlation \citep{chatterjee2021new, shi2022power}. These measures typically achieve their maximum value when \( Y \) is a measurable or monotone function of \( X \), rather than requiring \( X \) and \( Y \) to coincide almost surely.

The proposed profile association measure \eqref{d:indA} specifically targets the  quantification of  association   for random objects in general metric spaces. We note that  monotone functions are not well-defined in metric spaces that lack a natural order and furthermore that  alternative measures of independence often assign the same maximal association level to any monotone (or measurable) function of \( X \), providing no additional insight into the degree of dependence between \( Y \) and \( X \). This limitation prevents these measures from distinguishing varying levels of dependence.  In contrast, the proposed profile association uniquely identifies \( Y = X \) almost surely as the maximal form of dependence, providing a more precise interpretation of  maximal dependence. This property also indicates that test statistics based on \eqref{d:indA} may  have greater power in detecting dependence, and we find that  this is indeed the case  in the simulations reported in Section \ref{s:simu}.

\section{Methodology and theory}\label{s:meth}
\subsection{Testing independence for random objects}\label{s:test}
We are interested in testing
\[
\mathcal{H}_{0}: X \independent Y \quad \text{versus} \quad \mathcal{H}_{1}: X \nindependent Y
\]  
using random samples \(\mathcal{D}_{\mathsf{XY}}:= \{(X_i, Y_i) : 1 \leq i \leq n\}\) drawn from the joint distribution of the random objects \((X, Y)\).
The profile  association $\Delta(X,Y)$ in \eqref{d:indA}  quantifies the strength of the  relationship between random objects \(X\) and \(Y\) and provides the foundation
for the development of a test for   independence.

{As discussed in Section \ref{s:asso}, for scalar random variables $X$ and $Y$ with joint CDF $F$, \citet{hoeffding1948non} introduced the functional $
\Delta(F) = \iint \{F(x,y) - F(x,\infty)\,F(\infty,y)\}^2 \, \diff F(x,y)$ as a measure of dependence. 
This functional can be expressed in terms of indicator functions. 
Specifically, define $
\psi(x_1, x_2, x_3) = \mathds{1}_{\{x_1 - x_2 \geq 0\}} - \mathds{1}_{\{x_1 - x_3 \geq 0\}},$ where the indicator function $\mathds{1}_A = 1$ if the event $A$ is true and $\mathds{1}_A = 0$ otherwise. 
Setting  $
\phi(x_1, y_1; \ldots, x_5, y_5) = 0.25 \times \psi(x_1, x_2, x_3) \psi(x_1, x_4, x_5) \psi(y_1, y_2, y_3) \psi(y_1, y_4, y_5),$ one can write 
$$
\Delta(F) = \int \cdots \int \phi(x_1, y_1; \ldots, x_5, y_5)\, \diff F(x_1, y_1) \cdots \diff F(x_5, y_5).
$$
When $X$ and $Y$ are random objects, we define the scalar random variables $U_{\omega_1} = d_1(X, \omega_1)$ and $V_{\omega_2} = d_2(Y, \omega_2)$, and denote their joint CDF by $\mathfrak{F}_{\omega_1, \omega_2}$ for $\omega_1 \in \mathcal{M}_1$ and $\omega_2 \in \mathcal{M}_2$. Using these notations, the profile association defined in \eqref{d:indA} can be written as $
\Delta(X, Y) = \mathbb{E}_{X', Y'}\left[ \Delta\left( \mathfrak{F}_{X', Y'} \right) \right].
$}
 
Inspired by this, consider  the functions \(\psi_1: \mathcal{M}_1^{\otimes 4} \mapsto \{-1, 0, 1\}\)  defined as 
\[
\psi_1(x_1, x_2, x_3, x_4) = \mathds{1}_{\{d_1(x_1, x_3) \leq d_1(x_1, x_2)\}} - \mathds{1}_{\{d_1(x_1, x_4) \leq d_1(x_1, x_2)\}}, \quad x_i \in \mathcal{M}_1,
\]
and  \(\psi_2: \mathcal{M}_2^{\otimes 4} \mapsto \{-1, 0, 1\}\) defined as 
\[
\psi_2(y_1, y_2, y_3, y_4) = \mathds{1}_{\{d_2(y_1, y_3) \leq d_2(y_1, y_2)\}} - \mathds{1}_{\{d_2(y_1, y_4) \leq d_2(y_1, y_2)\}}, \quad y_i \in \mathcal{M}_2.
\]
For simplicity, we will omit the subscripts in \(\psi_1\), \(\psi_2\), $d_1$ and $d_2$, as the context will make it clear whether the functions are applied to elements in \(\mathcal{M}_1\) or \(\mathcal{M}_2\). We also define the function  
\begin{align*}
	h\big((x_{1},y_{1}),\ldots, (x_{6},y_{6})\big) =& \frac{1}{4} \psi(x_{1},x_{2},x_{3},x_{4}) \psi(x_{1},x_{2},x_{5},x_{6}) \\&\quad\times \psi(y_{1},y_{2},y_{3},y_{4}) \psi(y_{1},y_{2},y_{5},y_{6}).
\end{align*}

In what follows, we use the notation \(\{(X_1, Y_1), (X_2, Y_2), \ldots, (X_6, Y_6)\}\) to denote six independent copies of \((X, Y)\). Following the previous discussion, we have 
\[
\E\Bigl[h\bigl((X_{1},Y_{1}), \ldots, (X_{6},Y_{6})\bigr)\Bigr] \;=\; \Delta(X,Y).
\]
Since the function \( h \) is not symmetric, we first symmetrize it as follows, 
\[
\mathfrak{h}((x_{1},y_{1}),\ldots,(x_{6},y_{6}) ) = \frac{1}{6!} \sum_{\mathcal{P}_{6}^{6} (\bm{\alpha})} h((x_{\alpha_{1}},y_{\alpha_{1}}),\ldots,(x_{\alpha_{6}},y_{\alpha_{6}})),
\]
where \( \mathcal{P}_{6}^{6} (\bm{\alpha}) \) denotes the set of all permutations $\{\alpha_1,\ldots,\alpha_6\}$ of  \( \{1, 2, \ldots, 6\} \).
Given random samples $\mathcal{D}_{\mathsf{XY}}$, we employ a U-statistic as an estimator of $\Delta(X,Y)$, defined as 
\begin{equation}\label{d:Dn}
	D_{n}(\mathcal{D}_{\mathsf{XY}})= \binom{n}{6}^{-1} \sum_{\mathcal{C}_{6}^{n} (\bm{\alpha	}) }\mathfrak{h}\left((X_{\alpha_{1}},Y_{\alpha_{1}}),\ldots,(X_{\alpha_{6}},Y_{\alpha_{6}}) \right),
\end{equation} 
where  $\sum_{\mathcal{C}_{6}^{n} (\bm{\alpha	}) } $ demotes summation over all $\binom{n}{6}$ combinations of $6$ distinct elements $\{\alpha_1,\ldots,\alpha_6\}$ from $\{1,2,\ldots,n\}$.

We require some  concepts related to the analysis of U-statistics, which were first introduced by \cite{hoeffding1948non}, with further details provided in \cite{serfling2009approximation}. 
For \( r = 1, 2, \ldots, 6 \), let 
\begin{align*}
	&\mathfrak{h}_{r}((x_{1},y_{1}),\ldots,(x_{r}, y_{r}) )\\ =& \E\left[\mathfrak{h}((X_1,Y_1),\ldots,(X_{6},Y_{6}) ) \mid X_{1}=x_{1},Y_{1}=y_{1},\ldots,X_{r}=x_{r},Y_{r}=y_{r} \right]
\end{align*}
and note that  
\[
\E[\mathfrak{h}_{r}((X_{1},Y_{1}),\ldots,(X_{r}, Y_{r}) )] = \Delta(X,Y), \quad \text{for } r = 1, 2,\ldots, 6.
\]  
Define the centered version of \(\mathfrak{h}_{r}\) as  
\[
\tilde{\mathfrak{h}}_{r}((x_{1},y_{1}),\ldots,(x_{r}, y_{r}) ) = \mathfrak{h}_{r}((x_{1},y_{1}),\ldots,(x_{r}, y_{r}) ) - \Delta(X,Y),
\]  
and 
\[
\xi_{r} = \var\big[\tilde{\mathfrak{h}}_{r}((X_{1},Y_{1}),\ldots,(X_{r}, Y_{r}) ) \big], \quad r = 1, 2,\ldots, 6.
\]

The variance terms \(\xi_r\) play a key role in understanding the distributional properties of U-statistics, as they capture the variability of the centered kernels \(\tilde{\mathfrak{h}}_{r}\). Specifically, \(\xi_1\) corresponds to the variance of the first-order centered kernel and indicates whether the U-statistic has a non-degenerate linear term in its Hoeffding decomposition. When  \(\xi_1 = 0\) this is referred to as the degenerate case. This occurs when the first-order projection of the centered kernel function \(\tilde{\mathfrak{h}}_{1}\) is zero, indicating that the statistic is orthogonal to all linear functions of the data. Then  higher-order terms dominate the variance, often making the asymptotic distribution more complex. This situation often arises in independence testing, where the U-statistic is specifically designed to capture non-linear dependencies between random variables.

For \(x_1, x_2 \in \mathcal{M}_{1}\), consider  the functions \(g_{1}: \mathcal{M}_1 \times \mathcal{M}_1 \to [0,1]\),  defined as  
\begin{equation}\label{d:gx}
	g_1(x_1, x_2) := \E\left[\mathds{1}_{\{d(X, X') \leq \max\{d(x_1, X), d(x_2, X') \} \}}\right],
\end{equation}  
where \(X'\) is an independent copy of \(X\), and 
 \(\mathfrak{g}_{1}: \mathcal{M}_1 \times \mathcal{M}_1 \to \mathbb{R}\), defined as  
\begin{equation}\label{d:ggx}
	\mathfrak{g}_{1}(x_1, x_2) := g_1(x_1, x_2) - \E g_1(x_1, X) - \E g_1(X, x_2) + \E g_1(X', X).
\end{equation}  
By substituting \(x_1, x_2, X, X'\) in \eqref{d:gx} and \eqref{d:ggx} with \(y_1, y_2, Y, Y'\), one can similarly define \(g_2(y_1, y_2)\) and \(\mathfrak{g}_2(y_1, y_2)\), which map from \(\mathcal{M}_2 \times \mathcal{M}_2\) to \([0,1]\) and \(\mathbb{R}\), respectively.  
The following result  shows that the  test statistic for the profile independence test is a degenerate U-statistic and establishes its asymptotic distribution under \(\mathcal{H}_{0}\).

\begin{theorem}\label{thm:t1}
Under \(\mathcal{H}_{0}\), the statistic \(D_{n}(\mathcal{D}_{\mathsf{XY}})\), as defined in \eqref{d:Dn}, is a degenerate U-statistic with \(\xi_{1} = 0\) and  
\[
\xi_2 = \frac{1}{225} \E[\mathfrak{g}^2_1(X_1, X_2) \mathfrak{g}^2_2(Y_1, Y_2)].
\]  
Assuming that \(\E[\mathfrak{g}^2_1(X_1, X_2) \mathfrak{g}^2_2(Y_1, Y_2)] \neq 0\), we have  
\[
{n} (D_{n}(\mathcal{D}_{\mathsf{XY}}) - \Delta(X,Y)) \xrightarrow{d} {15}  \sum_{j=1}^{\infty} \lambda_{j}(Z_{j} - 1),
\]  
where \(Z_{j}\) are independent \(\chi_{1}^2\) random variables, and \(\lambda_{j}\) are the eigenvalues of the linear operator \(\mathscr{H} : \mathcal{L}_{\mu_{XY}}^{2}(\mathcal{M}_{1} \otimes \mathcal{M}_{2}) \to \mathcal{L}_{\mu_{XY}}^{2}(\mathcal{M}_{1} \otimes \mathcal{M}_{2})\), defined by  
\begin{align*}
\mathscr{H}(f) &= \iint \tilde{\mathfrak{h}}_{2}((x, y), (u, v))f(x, y) \diff \mu_{XY}(x, y), 
\end{align*}
for all $(u, v) \in \mathcal{M}_1 \times \mathcal{M}_2$ and $f \in \mathcal{L}_{\mu_{XY}}^{2}(\mathcal{M}_{1} \times \mathcal{M}_{2}) $.
\end{theorem}

\subsection{Permutation test and asymptotic results}\label{s:perm}

Since the asymptotic distribution in Theorem \ref{thm:t1} is a second-order Wiener chaos with unknown parameters, we use a permutation test to determine the critical values of the test statistic \(D_{n}(\mathcal{D}_{\mathsf{XY}})\) when implementing the test. 
Permuting the data and recalculating the test statistic for each permutation is a standard approach to generate a reference distribution. 
In this subsection, we implement the proposed independence test using a permutation-based method and study its asymptotic properties, including consistency and power analysis. 
Recently, alternative approaches such as the random-lifter method \citep{gao2025studentized} and spectrum-based methods \citep{wang:23:3} have been studied to handle independence testing in the presence of unknown limiting distributions. 
We include a discussion of the spectrum-based approach in the supplementary material.

For testing independence, a common  approach is to compute replicates of the test statistic by randomly permuting the \(Y\) sample while keeping the order of the \(X\) sample fixed \citep{szekely2007measuring}.
However, randomly permuting the entire \(Y\) sample introduces dependence among the data pairs under the alternative \(\mathcal{H}_1\) when $X$ and $Y$ are dependent, making it challenging to establish consistency of the critical value under \(\mathcal{H}_1\).
To address this issue, we adopt a half-permutation procedure introduced in  \cite{fan2024test} that does not require a splitting technique with its possible loss of power. Adopting this approach for our setting requires  a different   proof technique to obtain consistency and power bounds for the proposed profile independence test, as  the previous techniques are not applicable. 

We use $\mathcal{P}_{\lfloor n/2 \rfloor }^{n}$ to denote the set of all subsets of $\{1, 2, \ldots, n\}$ of size $\lfloor n/2 \rfloor$, where each subset consists of unrepeated random samples from $\{1, \ldots, n\}$. Consider a sequence of i.i.d. random permutations \(\Pi_{1}, \ldots, \Pi_{B}\) that are independent of \((X, Y)\) and uniformly distributed on \(\mathcal{P}_{\lfloor n/2 \rfloor }^{n}\). We use \(\Pi_{j}^{c}\) to denote the complement of \(\Pi_{j}\) with respect to the set \(\{1, 2, \ldots, n\}\). Without loss of generality, we assume \(n\) is even, so that \(\Pi_{j}\) and \(\Pi_{j}^{c}\) have the same cardinality \(n/2\). For each \(\Pi_{j}^{c}\), let \(\{\sigma_{jk}\}_{k=1}^{N_j}\) be a sequence of random permutations on \(\Pi_{j}^{c}\) that are independent of \((X, Y)\) and \(\Pi_{j}\).
Assume \(\Pi_{j} = \{\imath_1, \imath_2, \ldots, \imath_{\lfloor n/2 \rfloor}\}\) and \(\sigma_{jk} \circ \Pi_{j}^{c} = \{\jmath_1, \jmath_2, \ldots, \jmath_{\lfloor n/2 \rfloor}\}\), where \(\sigma_{jk} \circ \Pi_{j}^{c}\) denotes the sequence of \(\Pi_{j}^{c}\) under the permutation \(\sigma_{jk}\). The permuted data corresponding to \((\Pi_{j}, \sigma_{jk})\) are
\[
\mathcal{D}_{\mathsf{XY}}(\Pi_{j}, \sigma_{jk}) := \{(X_{\imath_1}, Y_{\jmath_1}), (X_{\imath_2}, Y_{\jmath_2}), \ldots, (X_{\imath_{\lfloor n/2 \rfloor}}, Y_{\jmath_{\lfloor n/2 \rfloor}})\},
\] and 
the test statistic  based on permuted data $\mathcal{D}_{\mathsf{XY}}(\Pi_{j}, \sigma_{jk}) $ is  
$$D_{\lfloor n/2\rfloor} (\mathcal{D}_{\mathsf{XY}}(\Pi_{j}, \sigma_{jk})):= \binom{\lfloor n/2 \rfloor }{6}^{-1} \sum_{\mathcal{C}_{6}^{\lfloor n/2 \rfloor} (\bm{\alpha})} h((X_{\imath_{ \alpha_{1}}}, Y_{\jmath_{ \alpha_{1}}}), \ldots,(X_{\imath_{ \alpha_{6}}}, Y_{\jmath_{ \alpha_{6}}})). $$

Let \(N = \sum_{j=1}^{B} N_j\) be the total number of permutation pairs \((\Pi_{j}, \sigma_{jk})\). Since the pairs \((\Pi_{j}, \sigma_{jk})\) are random elements, we can define a randomized test at level \(\alpha\) by  
{\begin{equation}\label{d:rt}
	\psi_{\alpha}(\mathcal{D}_{\mathsf{XY}} ) :=\mathds{1}\left(1 + \sum_{(\Pi_{j}, \sigma_{jk})} \mathds{1}_{\{n\mathcal{D}_{n}(\mathcal{D}_{\mathsf{XY}}) \leq \lfloor n/2\rfloor D_{\lfloor n/2\rfloor} (\mathcal{D}_{\mathsf{XY}}(\Pi_{j}, \sigma_{jk}))\}} \leq (1 + N)\alpha \right),
\end{equation} 
where 
the only randomness in $\psi_{\alpha}(\mathcal{D}_{\mathsf{XY}})$  comes from the random permutations $(\Pi_{j}, \sigma_{jk})$ and $\psi_{\alpha}(\mathcal{D}_{\mathsf{XY}} ) =1$  indicates that the test rejects $H_0$ at level $\alpha$. }

Under the null hypothesis, \(X\) and \(Y\) are independent, and the elements in the data sets \(\mathcal{D}_{\mathsf{XY}}\) and \(\mathcal{D}_{\mathsf{XY}}(\Pi_{j}, \sigma_{jk})\) are exchangeable. Consequently, \(\{\lfloor n/2\rfloor D_{\lfloor n/2\rfloor} (\mathcal{D}_{\mathsf{XY}}(\Pi_{j}, \sigma_{jk}))\}_{j,k}\) and \(n\mathcal{D}_{n}(\mathcal{D}_{\mathsf{XY}})\) have the same asymptotic distribution given by Theorem \ref{thm:t1}. Therefore, the probability on the right-hand side of equation \eqref{d:rt} equals \(\alpha\) for all distributions of \((X, Y)\) under the null hypothesis, which implies that \(\psi_{\alpha}(\mathcal{D}_{\mathsf{XY}} )\) is a valid size-\(\alpha\) test. For the consistency under the alternative $\mathcal{H}_1$, denote by \(H(z)\)  the cumulative distribution function of the asymptotic null distribution in Theorem \ref{thm:t1}. We employ the estimator 
\begin{equation}\label{d:hatH}
	\hat{H}_{N}(z) = \frac{1}{N} \sum_{(\Pi_{j}, \sigma_{jk})} \mathds{1}_{\left\{\lfloor \frac{n}{2} \rfloor D_{\lfloor n/2 \rfloor}( \mathcal{D}_{\mathsf{XY}}(\Pi_{j}, \sigma_{jk})) \leq z \right\}}
\end{equation}
for  \(H(z)\) based on the permuted data. Theorem \ref{thm:pcon} below establishes the consistency of this estimator, which ensures  that the size of the proposed profile independence test is correctly controlled under the alternative.
\begin{theorem}\label{thm:pcon}
	Under the conditions of Theorem \ref{thm:t1} and if the  number of permutations satisfies  $N\rightarrow\infty$, it holds that $$|\hat{H}_{N}(z)-H(z)|=o_{P}(1)$$  for every $z$ which is a continuity point of $H(z)$.
\end{theorem}

To further investigate the asymptotic power behavior of \eqref{d:rt}, we consider the alternative space \(\mathcal{F}_{1}(\rho) := \{ \mu_{XY} \in \mathcal{F} : \Delta(X, Y) \geq \rho\}\) for any positive \(\rho\), where \(\mathcal{F}\) denotes the subset of all probability measures on \(\mathcal{M}_{1} \times \mathcal{M}_2\) whose cumulative distribution functions satisfy the assumptions in Theorem \ref{thm:assD}. Since we have already shown that the type-I error of ${\psi}_{\alpha}(\mathcal{D}_{\mathsf{XY}})$ is always bounded by $\alpha$, we  define the minimax separation radius as
\begin{equation}\label{df:sep}
	\rho^{\ast}(n,\alpha, \beta) = \inf\left\{\rho > 0: \, \inf_{{\psi} \in \mathsf{\Psi}(\alpha)} \sup_{\mu_{XY} \in \mathcal{F}_{1}(\rho)} \E(1-\psi)   \leq \beta \right\},
\end{equation}
where $\mathsf{\Psi}(\alpha)$ denotes the set containing all valid size-$\alpha$ independence tests and the expectation 
is taken over the 
joint probability measure $\mu_{XY}$. The separation \eqref{df:sep} quantifies the smallest difference between the null and alternative hypotheses in terms of 
$ \Delta(X, Y)$ for which the  test can reliably distinguish between null and alternative under specified  upper bounds on the  type-I and type-II errors. The following result  provides an  upper bound for the separation distance $\rho^{\ast}(n,\alpha, \beta)$. 
\begin{theorem}\label{thm:sep}
	For any fixed $\alpha, \beta \in (0,1)$ such that $\alpha + \beta < 1$, and assuming the number of permutations $N$ satisfies $N > (\alpha\beta)^{-1} - 1$, there exists a constant $C$ that depends only on $\alpha$ and $\beta$, such that
$$\rho^{\ast}(n,\alpha, \beta) \leq C n^{-1/2}.$$
\end{theorem}

\section{Conditional independence test}\label{s:cond}

Conditional  independence testing  is a statistical tool used to determine whether two variables are independent when conditioning on  a third variable. 
It is used in a wide range of statistical applications where one studies complex relationships among variables, including causal inference \citep{zhang2011kernel, zhang2017causal, peters2017elements}, graphical models \citep{koller2009probabilistic, williams2020bayesian} and sufficient dimension reduction \citep{cook2004testing, chen2015diagnostic, li2018sufficient}.
Many nonparametric conditional independence tests have been studied based on conditional density functions \citep{huang2010testing, su2008nonparametric, banerjee2024ball}, distribution functions \citep{linton1996conditional, zhou2020test} and characteristic functions \citep{su2007consistent, wang2015conditional, wang2018characteristic}. These methods are all restricted to the case where data reside in a vector space. To our knowledge, 
conditional independence tests for random elements in general metric spaces that do not support linear operations have remained unexplored. 

We consider a setting where \( X \in \mathcal{M}_1 \) and \( Y \in \mathcal{M}_2 \) are random objects, while the conditional variable $Z$ is Euclidean, specifically $Z \in \mathcal{Z} $, where $ \mathcal{Z}$ is a compact subset of $\reall^p$. Our goal is to test the following hypothesis:
\begin{equation}\label{d:cit}
	\mathcal{H}_{0,c}: X \mid Z \independent Y \mid Z \quad \text{versus} \quad \mathcal{H}_{1,c}: X \mid Z \nindependent Y \mid Z,
\end{equation}
using i.i.d. samples \(\{(X_i, Y_i, Z_i)\}_{i=1}^n\) drawn from the joint distribution of \((X, Y, Z)\). 
To develop a suitable test statistic, we  extend the concepts of marginal and joint distribution profiles to their conditional counterparts. For each $z\in\mathcal{Z}$, the marginal and joint conditional distribution profiles for \( X \) and \( Y \) are defined as 
\[
\begin{aligned}
	&F_{\omega_1, z}^X(u) = \mathbb{P}\big(d(X, \omega_1) \leq u \mid Z = z \big), \quad \text{ for all } \omega_1 \in \mathcal{M}_1, \, u \geq 0, \\
	&F_{\omega_2, z}^Y(v) = \mathbb{P}\big(d(Y, \omega_2) \leq v \mid Z = z \big), \quad \text{ for all } \omega_2 \in \mathcal{M}_2, \, v \geq 0;
\end{aligned}
\]
and
\begin{align*} 
F_{\omega_1, \omega_2, z}^{XY}(u, v) = &\mathbb{P}\big(d(X, \omega_1) \leq u, \, d(Y, \omega_2) \leq v \mid Z = z \big)
\end{align*}
for all $ \omega_1 \in \mathcal{M}_1, \, \omega_2 \in \mathcal{M}_2, \, u, v \geq 0 $.

Based on   these conditional  distance profiles, we define the \emph{conditional profile association} 
\begin{equation}\label{d:cpda0}
	\Delta_{XY}(z)=\mathbb{E}\left[ \iint \left\{F_{X',Y',Z'}^{XY}(u,v) - F_{X',Z'}^{X}(u)F_{Y',Z'}^{Y}(v)\right\}^2 \diff F_{X',Y',Z'}^{XY}(u,v) \Bigm|Z'=z \right],
\end{equation}  
where $F_{X',Y',Z'}^{XY}$, $F_{X',Z'}^{X}$ and $F_{Y',Z'}^{Y}$ are given by
\[
\begin{aligned}
    &F_{X',Y',Z'}^{XY}(u,v) := \mathbb{P}_{X,Y,Z}\left(d(X, X') \leq u, \, d(Y, Y') \leq v\mid Z=Z' \right), \\
    &F_{X',Z'}^{X}(u) := \mathbb{P}_{X,Z}\left(d(X, X') \leq u \mid Z=Z'\right), \, F_{Y',Z'}^{Y}(v) := \mathbb{P}_{Y,Z}\left(d(Y, Y') \leq v \mid Z=Z' \right).
\end{aligned}
\]
We use \(\Delta_{XY}(z)\) as  measure of conditional profile  association between \(X\) and \(Y\) given \(Z=z\).  
When \(Z\) is a continuous random variable, such as time, conditional  profile association can  be used to quantify the time-varying behavior of the dependency  between \(X\) and \(Y\) as a function of  \(Z\). Such measures have  important applications in many areas, including  the life and social sciences and climatology. We  provide a specific application example for the analysis of human longevity in Section \ref{s:data}. 

The definition \eqref{d:cpda0} places no restrictions on the distribution of $Z$; hence, it remains valid whether $Z$ is continuous, discrete, or of mixed type. 
We focus on the case where $Z$ is univariate and continuous in the estimation procedure, as this setting is both commonly encountered in practice and theoretically nontrivial, particularly due to the challenges involved in establishing uniform convergence for local linear smoothing with object-valued data. 
When $Z$ is discrete (e.g., categorical), the estimation becomes significantly simpler: one can compute the association measure within each category of $Z$ and aggregate the results, avoiding the need for smoothing altogether.
To avoid the curse of dimensionality under some structural assumptions, for the  multivariate case one can apply 
single index  Fr\'echet regression \citep{bhattacharjee2023single},  employing   arguments that are similar to those  in  \cite{zhou2024conformal} for the case of multivariate predictors in conformal inference. 
For the case of a univatiate continuous predictor $Z$ we use nonparametric regression with argument $Z$ to implement conditional profiles, employing   local linear estimators \citep{fan1992variable} for \( F_{\omega_1,z}^{X}(u) \), \( F_{\omega_2,z}^{Y}(v) \), and \( F_{\omega_1,\omega_2,z}^{XY}(u,v) \). This yields  
\begin{align*}
	\hat{F}^{X}_{\omega_1, z}(u) =& \argmin{\beta_{0} \in \mathbb{R}} \frac{1}{n h_n} \sum_{i=1}^{n} \left\{\mathds{1}_{\{d(X_i,\omega_1) \leq u \}} - \beta_{0} - \beta_{1}(Z_i - z) \right\}^2 \mathrm{K}\left(\frac{Z_i - z}{h_n}\right);\\
	\hat{F}^{Y}_{\omega_2, z}(v) =& \argmin{\beta_{0} \in \mathbb{R}} \frac{1}{n h_n} \sum_{i=1}^{n} \left\{\mathds{1}_{\{d(Y_i,\omega_2) \leq v \}} - \beta_{0} - \beta_{1}(Z_i - z) \right\}^2 \mathrm{K}\left(\frac{Z_i - z}{h_n}\right);\\
	\hat{F}^{XY}_{\omega_1,\omega_2,z}(u,v) = &\argmin{\beta_{0} \in \mathbb{R}} \frac{1}{n h_n} \sum_{i=1}^{n} \left\{ \mathds{1}_{\{d(X_i,\omega_1) \leq u \}} \mathds{1}_{\{d(Y_i,\omega_2) \leq v \}} - \beta_{0} - \beta_{1}(Z_i - z) \right\}^2\\
	&\quad\quad\quad\quad \times \mathrm{K}\left(\frac{Z_i - z}{h_n}\right),  \hspace{13cm}
\end{align*}
where \(\mathrm{K}(\cdot)\) is a symmetric and Lipschitz continuous density kernel on \([-1,1]\)  and \(h_n\) is a  sequence of bandwidths. 

To study the asymptotic behavior of these conditional distance profile  estimators, we introduce the sets of functions  
\begin{align*} \mathcal{F}_{1}&=\,\{f(x)=\mathds{1}_{\{d(\omega_1,x)\leq t \}}:\omega_1\in\mathcal{M}_1\,\,\,   t\in \reall^{+} \},  \\  \mathcal{F}_{2}&=\,\{f(y)=\mathds{1}_{\{d(\omega_2,y)\leq t \}}:\omega_2\in\mathcal{M}_2,\,\,\,  t\in \reall^{+} \},\end{align*}   which are  classes of indicator functions indexed by $\omega_1$, $\omega_2$ and  $t$.
For a bounded function class \(\mathcal{F}\) defined on \(\mathcal{M}\), the  uniform covering number 
is $\mathcal{N}(\epsilon, \mathcal{F}) := \sup_{\mathbb{Q}} N(\epsilon, \mathcal{F}, d_{\mathbb{Q}}),
$
where the supremum is taken over all probability measures \(\mathbb{Q}\) on \(\mathcal{M}\) with bounded second-order moments. Here, \(d_{\mathbb{Q}}\) is the \(\mathcal{L}_{\mathbb{Q}}^2\) metric, defined for any two functions \(f, g \in \mathcal{F}\) as:
$
d_{\mathbb{Q}}^2(f, g) = \int \{f(x) - g(x)\}^2 \, \mathrm{d}\mathbb{Q}(x).
$ We require  the following regularity assumptions. 


\begin{assumption}\label{as:a1}
	The marginal distribution of $Z$ has a continuous density function $f_Z$, which satisfies $\inf_{z\in\operatorname{Support}(f_{Z})} f_Z(z)>c_1$ and $\sup_{z\in\tdomain} f_Z(z)<c_{2}$ for  strict positive constants $c_1$ and $c_2$.
\end{assumption}
\begin{assumption}\label{as:a2}
	The bandwidth sequence $\{h_{n}\}_{n\geq 1}$ satisfies $nh_{n}/\log n\rightarrow\infty$  and $|\log h_n|/\log\log n\rightarrow\infty$ as $n\rightarrow\infty$.
\end{assumption}

\begin{assumption}\label{as:a3}
	For every $\omega_1 \in \mathcal{M}$, $\omega_2\in\mathcal{M}_2$ and $z\in \tdomain$,  the distance  profile $F^{XY}_{\omega_1,\omega_2, z}$ is absolutely continuous with continuous density $f_{\omega_1,\omega_2, z}$ and there exist strict positive constants $c_{3}$ and $c_{4}$ such that $\inf _{t \in \operatorname{support}\left(f_{\omega_1,\omega_2, z}\right)} f_{\omega_1,\omega_2, z}(t)\geq c_3 $ and $\sup _{t \in \mathbb{R}^{+}} f_{\omega_1,\omega_2, z}(t)\leq c_{4}<\infty$. 
\end{assumption}
\begin{assumption}\label{as:a4}
	For every $\omega_1 \in \mathcal{M}_1$, $\omega_2 \in \mathcal{M}_2$ and $u,v\in \reall^{+}$,  $F^{XY}_{\omega_1,\omega_2, z}(u,v)$ is second order differentiable and has bounded second order derivatives with respect to $z$. Moreover, there exists a constant $L'$ such that $ |F^{XY}_{\omega_1,\omega_2, z}(u,v)-F^{XY}_{\omega_1',\omega_2, z}(u,v)|\leq L'd_1(\omega_{1},\omega_{1}')$, $ |F^{XY}_{\omega_1,\omega_2, z}(u,v)-F^{XY}_{\omega_1,\omega_2', z}(u,v)|\leq L'd_2(\omega_{2},\omega_{2}')$ for all $z\in\tdomain,u,v\in\reall^{+}$ and $\omega_{1},\omega_{1}',\in \mathcal{M}_1$, $\omega_{2},\omega_{2}',\in \mathcal{M}_2$.
\end{assumption}

\begin{prop}\label{p:locl} 
	Under assumptions \ref{as:a1} and \ref{as:a4}, if \(\max\{\mathcal{N}(\epsilon,\mathcal{F}_1),\, \mathcal{N}(\epsilon,\mathcal{F}_2)\} \lesssim \epsilon^{-v}\) for some \(v > 0\) and \(h_n \asymp (n / \log n)^{-1/5}\), 
$$
\begin{aligned}
	&\sup_{\omega_1 \in \mathcal{M}_1} \sup_{z \in \mathcal{Z}} \sup_{u \geq 0} \left|\hat{F}^{X}_{\omega_1, z}(u) - F^{X}_{\omega_1, z}(u)\right| = O\left(\left( \frac{\log n}{ n}\right)^{\frac{2}{5}} \right) \text{ a.s.}, \\
	&\sup_{\omega_2 \in \mathcal{M}_2} \sup_{z \in \mathcal{Z}} \sup_{v \geq 0} \left|\hat{F}^{Y}_{\omega_2, z}(v) - F^{Y}_{\omega_2, z}(v)\right| = O\left(\left( \frac{\log n}{ n}\right)^{\frac{2}{5}} \right) \text{ a.s.}, \\
	&\sup_{\omega_1 \in \mathcal{M}_1} \sup_{\omega_2 \in \mathcal{M}_2} \sup_{z \in \mathcal{Z}} \sup_{u,v \geq 0} \left|\hat{F}^{XY}_{\omega_1, \omega_2, z}(u,v) - F^{XY}_{\omega_1, \omega_2, z}(u,v)\right| = O\left(\left( \frac{\log n}{ n}\right)^{\frac{2}{5}} \right) \text{ a.s.}.
\end{aligned}
$$
\end{prop}

An estimator for the  conditional profile  association ${\Delta}_{XY}(z)$ is then obtained through another nonparametric regression step, 
\begin{equation}\label{eq:cpdahat}
	\hat{\Delta}_{XY}(z)=\argmin{\beta_{0} \in \mathbb{R}} \frac{1}{n h_n} \sum_{j=1}^{n} \left\{\hat{R}_j - \beta_{0} - \beta_{1}(Z_j - z) \right\}^2 \mathrm{K}\left(\frac{Z_j - z}{h_n}\right), 
\end{equation} 
where 
$$\hat{R}_{j}=\iint \left\{\hat{F}_{X_j,Y_j,Z_j}^{XY}(u,v) - \hat{F}_{X_j,Z_j}^{X}(u)\hat{F}_{Y_j,Z_j}^{Y}(v)\right\}^2 \diff u\diff v. $$
The following result  provides the uniform convergence of this estimator.

\begin{theorem}\label{thm:cpda}
	Under the assumptions of Proposition \ref{p:locl}, 
	$$\sup_{z\in\mathcal{Z}} |\hat{\Delta}_{XY}(z)-{\Delta}_{XY}(z) |=O\left( h^2+\sqrt{\frac{\log n}{nh_n}} \right) \text{ a.s.}. $$
\end{theorem}
This result provides theoretical justification for the estimator \eqref{eq:cpdahat} and supports its use as a measure for the conditional 
association of random effects. 
Using  the estimated conditional profiles we then obtain the following  test statistic for testing \eqref{d:cit},
\begin{align} \label{Tn}
\mathcal{T}_{n} = \frac{1}{n} \sum_{i=1}^{n} \iint \left\{\hat{F}^{XY}_{X_i,Y_i,Z_i}(u,v) - \hat{F}^{X}_{X_i,Z_i}(u) \hat{F}^{Y}_{Y_i,Z_i}(v) \right\}^2 \diff \hat{F}_{X_i,Y_i,Z_i}^{XY}(u,v).
\end{align} 

To study the  asymptotic distribution of $\mathcal{T}_{n}$, we define the following covariance operator 
\begin{align} \label{covop}
		\mathcal{C}_{x,y,z}(u_1,u_2,v_1,v_2)\,=\, &C^{XY}_{x,y,z}(u_1,u_2,v_1,v_2)+F^{X}_{x,z}(u_1)F^{X}_{x,z}(u_2)C^{Y}_{y,z}(v_1,v_2)\\& \hspace{-3cm} + F^{Y}_{y,z}(v_1)F^{Y}_{y,z}(v_2)C^{X}_{x,z}(u_1,u_2)+\mathcal{C}_{1}(u_1,u_2,v_1,v_2) +\mathcal{C}_{2}(u_1,u_2,v_1,v_2), \nonumber
	\end{align}
where
\begin{align*}
    C^{XY}_{x,y,z}(u_1,u_2,v_1,v_2) &= 
    \mathrm{Cov}\left( \frac{1}{f_{Z}(z)} \mathrm{K}\left(\frac{Z-z}{h}\right)\mathds{1}_{\{d(x,X) \leq u_1\}} \mathds{1}_{\{d(y,Y) \leq v_1\}}, \right. \\
    &\qquad\qquad \left. \frac{1}{f_{Z}(z)}\mathrm{K}\left(\frac{Z-z}{h}\right)\mathds{1}_{\{d(x,X) \leq u_2\}} \mathds{1}_{\{d(y,Y) \leq v_2\}}\right);\\
    C^{X}_{x,z}(u_1,u_2) &=  \mathrm{Cov}\left(\frac{1}{f_{Z}(z)} \mathrm{K}\left(\frac{Z-z}{h}\right)\mathds{1}_{\{d(x,X) \leq u_1\}}, \right. \\
    &\qquad\qquad \left. \frac{1}{f_{Z}(z)} \mathrm{K}\left(\frac{Z-z}{h}\right)\mathds{1}_{\{d(x,X) \leq u_2\}}\right);\\
    C^{Y}_{y,z}(v_1,v_2) &=  \mathrm{Cov}\left(\frac{1}{f_{Z}(z)} \mathrm{K}\left(\frac{Z-z}{h}\right)\mathds{1}_{\{d(y,Y) \leq v_1\}}, \right. \\
    &\qquad\qquad \left. \frac{1}{f_{Z}(z)} \mathrm{K}\left(\frac{Z-z}{h}\right)\mathds{1}_{\{d(y,Y) \leq v_2\}}\right).\\
    	\mathcal{C}_{1}(u_1,u_2,v_1,v_2)&= \mathrm{Cov}\left( \frac{1}{f_{Z}(z)} \mathrm{K}\left(\frac{Z-z}{h}\right)\mathds{1}_{\{d(x,X) \leq u_1\}} \mathds{1}_{\{d(y,Y) \leq v_1\}}, \right. \\
    &\qquad\qquad \left. F_{x,z}^{X}(u_2) \frac{1}{f_{Z}(z)}\mathrm{K}\left(\frac{Z-z}{h}\right) \mathds{1}_{\{d(y,Y) \leq v_2\}}\right)\\
    \mathcal{C}_{2}(u_1,u_2,v_1,v_2)&= \mathrm{Cov}\left( \frac{1}{f_{Z}(z)} \mathrm{K}\left(\frac{Z-z}{h}\right)\mathds{1}_{\{d(x,X) \leq u_1\}} \mathds{1}_{\{d(y,Y) \leq v_1\}}, \right. \\
    &\qquad\qquad \left. F_{y,z}^{Y}(v_2) \frac{1}{f_{Z}(z)}\mathrm{K}\left(\frac{Z-z}{h}\right) \mathds{1}_{\{d(x,X) \leq u_2\}}\right).
\end{align*}


\begin{theorem}\label{thm:ctn}
	Under  Assumptions \ref{as:a1} to \ref{as:a4}, $nh^2\log n\rightarrow 0 $ and $\mathcal{H}_{0,c}$, \( nh\mathcal{T}_n \) converges in distribution to the law of a random variable \( L = \sum_{k=1}^{\infty} Z_k^2 \mathbb{E}_{X,Y,Z}(\lambda_k^{X,Y,Z}) \), where \( Z_1, Z_2, \ldots \) is a sequence of i.i.d. \( \mathcal{N}(0,1) \) random variables, and \( \lambda_1^{x,y,z} \geq \lambda_2^{x,y,z} \geq \cdots \) are the eigenvalues of the covariance operator \eqref{covop}.
\end{theorem}

Since the limit distribution  in the  theorem above is unknown, we implement the test using a permutation approach as introduced in Section \ref{s:perm}. The permuted data are
\[
\mathcal{D}_{\mathsf{XYZ}}(\Pi_{j}, \sigma_{jk}) := \{(X_{\imath_1}, Y_{\jmath_1}, Z_{\jmath_1}), (X_{\imath_2}, Y_{\jmath_2}, Z_{\jmath_2}), \ldots, (X_{\imath_{\lfloor n/2 \rfloor}}, Y_{\jmath_{\lfloor n/2 \rfloor}}, Z_{\jmath_{\lfloor n/2 \rfloor}})\}.
\]
This permutation procedure is similar to  that presented in \cite{berrett2020conditional}.
Let $\mathcal{T}_{\lfloor n/2 \rfloor }(\mathcal{D}_{\mathsf{XYZ}}(\Pi_{j}, \sigma_{jk}))$ represent the test statistic defined in \eqref{Tn} based on the permuted data $\mathcal{D}_{\mathsf{XYZ}}(\Pi_{j}, \sigma_{jk})$, and let $\mathcal{T}_{n}(\mathcal{D}_{\mathsf{XYZ}})$ be the corresponding test statistic calculated using the original data.
The $p$-value resulting from the test is then 
\[
p_{c} := \frac{1 + \sum_{\Pi_{j}, \sigma_{jk}} \mathds{1}_{\{\lfloor n/2 \rfloor\mathcal{T}_{\lfloor n/2 \rfloor }(\mathcal{D}_{\mathsf{XYZ}}(\Pi_{j}, \sigma_{jk})) \geq n\mathcal{T}_{n}(\mathcal{D}_{\mathsf{XYZ}})\}}}{1 + N},
\]
where $N$ denotes the total number of permutations. The randomized test for testing \eqref{d:cit} at significance level $\alpha$ is then defined as {$\psi_{\alpha}^{c} :=\mathds{1}(p_{c}\leq \alpha)$.} 

Denoting  by $\Gamma(z)$ the cumulative distribution  function of    the limiting distribution in Theorem \ref{thm:ctn}, its corresponding  permutation-based estimator is 
\[
\hat{\Gamma}_{N}(z) = \frac{1}{N} \sum_{(\Pi_{j}, \sigma_{jk})} \mathds{1}_{\left\{\frac{n}{2}\mathcal{T}_{\lfloor n/2 \rfloor }( \mathcal{D}_{\mathsf{XYZ}}(\Pi_{j}, \sigma_{jk})) \leq z \right\}}, 
\]
for which we obtain the following convergence result.
\begin{corollary}\label{prop:ct}
	Under the same assumptions as in Theorem \ref{thm:ctn}, $|\hat{\Gamma}_{N}(z)-\Gamma(z)| =o_{P}(1)$ holds for every continuity point $z$ of $\Gamma(z)$.\end{corollary}

\section{{Simulations}}\label{s:simu}
We present finite-sample comparisons of the proposed test against several alternative methods when $X$ and $Y$ are located in different metric spaces. These include the Euclidean space $\mathbb{R}^{p}$, the sphere $\mathbb{S}^p$, the space of $d$-dimensional symmetric positive definite matrices $\mathcal{S}_{p}^{++}$, and the 2-Wasserstein space of one-dimensional distributions.
For the Euclidean space $\mathbb{R}^{p}$ and the sphere $\mathbb{S}^p$, we use the usual $\mathcal{L}^2$ distance and the geodesic metric, respectively. For the $\mathcal{S}_{p}^{++}$ space, we use the affine-invariant Riemannian distance $d(A,B)=\|\mathfrak{log}(A^{-1}B)\|_{F}^2$, where $\mathfrak{log}$ and $\mathfrak{exp}$ are the logarithm and exponential functions for matrices. Moreover, for $A,B\in \mathcal{S}_{p}^{++}$, the affine-invariant geodesic interpolation is defined by
$$A\#_{\rho}B=A^{1/2}(A^{-1/2}BA^{-1/2})^{\rho}A^{1/2}.$$
For the Wasserstein space \(\mathbb{W}^2\), we use the 2-Wasserstein metric, which is defined as 
 \begin{equation}  \label{wass} d_W(x,y)^2=\int_0^1  \{Q_x(u)- Q_y(u)\}^2 \, \diff u. \end{equation}
Here  $Q_x,\, Q_y$ are the quantile functions associated with the distributions $x$ and $y$. 
For each setting, both the proposed and comparison methods were implemented using the permutation-based test and empirical power curves at significance level  0.05 are obtained from  400 Monte Carlo runs.

For the unconditional testing problem in Section \ref{s:meth}, we consider the  following settings,  where in all settings the dependence is quantified by the parameter $\rho$, where $\rho=0$ corresponds to independence and the level of dependence (association) increases with $\rho$. We obtained empirical  power curves that illustrate how the power of the  profile independence test increases as the argument $\rho$ increases, where we consider  $0 \le\rho\le 1.$

\begin{itemize}
	\item  \textbf{$\mathbb{R}^{p}$-Lin}: $\bm{X} = (X_1, \ldots, X_p)^\top$ and $\bm{Y} = (Y_1, \ldots, Y_p)^\top$  are related only through their first $J_p = \max\{1, \lfloor 0.1p \rfloor\}$ components, where each $Y_j$ for $j \leq J_p$ is generated as $Y_j = \rho X_j + \varepsilon_j,$ with $\varepsilon_j \sim \mathcal{N}(0,1)$ independent of all other variables. The remaining components $Y_{J_p+1}, \ldots, Y_p$, and $X_1, \ldots, X_p$, are independently distributed as $\mathcal{N}(0,1)$.
	\item  \textbf{$\mathbb{R}^{p}$-Log}: $\bm{X} = (X_1, \ldots, X_p)^\top$ and $\bm{Y} = (Y_1, \ldots, Y_p)^\top$ are generated as in the $\mathbb{R}^p$-Lin setting, except that $Y_j = \rho \log(4X_j^2) + \varepsilon_j$, where $\varepsilon_j \sim \mathcal{N}(0,1)$ is independent of all other variables.
	\item \textbf{\(\mathbb{R}^{1}\)-Cir}: Denoting by \(\|v\|_{\zeta}$ the $\zeta$-norm for any Euclidean vector $v$. For any $1 \le  \zeta \le \infty$ we consider   pairs \((X, Y)\) of one-dimensional random variables $X$ and $Y$ that satisfy the condition \(\|(X, Y)\|_{2/\rho} \leq 1\), where \(X\) and \(Y\) are independently drawn from \(\text{Unif}(-1,1)\). When \(\rho = 0\), \((X, Y)\) is uniformly distributed over the square \([-1,1] \times [-1,1]\). When \(\rho = 1\), \((X, Y)\) follows a uniform distribution within the unit circle \(\{(x, y) \mid x^2 + y^2 \leq 1\}\), with domains for the uniform distribution located in between the square and the circle for  $0<\rho<1.$ 
	\item \textbf{$\mathcal{S}_{2}^{++}$-Interpolation}:  $X$ and $Y$ are $2\times2$ symmetric positive definite matrices.  $X$ is generated as $X=\mathfrak{exp}((S+S^{\top})/2)$, where $S$ is a $2\times 2$ matrix with independent entries drawn from $\mathcal{N}(0,0.6^2)$. Let $Y_{(0)}$ denote the independent baseline, which is generated independently but follows the same distribution as $X$. The dependent target $Y_{(1)}$ incorporates geodesic noise with $P=CXC^{\top}$, and is constructed as
$$Y_{(1)}=P^{1/2}\mathfrak{exp}(0.1(E+E^{\top})/2)P^{1/2},$$
where $C=\diag(2,0.5)$ and $E$ is a $2\times 2$ matrix with independent standard normal entries. The final matrix $Y$ is obtained via geodesic interpolation between the independent baseline $Y_{(0)}$ and the dependent target $Y_{(1)}$, defined as
$$Y=Y_{(0)}\#_{\rho/5}Y_{(1)}=Y_{(0)}^{1/2}(Y_{(0)}^{-1/2}Y_{(1)}Y_{(0)}^{-1/2})^{\rho/5}Y_{(0)}^{1/2}.$$

	\item \textbf{$\mathbb{S}^2$\&$\mathcal{S}_{2}^{++}$-Hybrid}: Here we consider $X \in \mathbb{S}^2$ as  on the unit sphere, while $Y \in \mathcal{S}_{2}^{++}$ is a symmetric positive definite matrix. First, generate $X_{(0)}$ and $Y$ as in the setting \textbf{$\mathcal{S}_{2}^{++}$-Interpolation}. Then, $X$ is the vectorized form of the upper triangular part of $X_{(0)}$, normalized so that $\|X\|_2=1$.

	\item \textbf{$\mathbb{W}^2$-Mean}: \(X\) and \(Y\) are random distributions in \(\mathbb{W}^2\), specifically  \(X =\mathcal{N}(\mu_{X},\sigma_{X})\) and \(Y=\mathcal{N}(\mu_{Y},\sigma_{Y})\), where \(\mu_{X} \sim \mathcal{N}(0,1) \),  \(\mu_{Y}=0.2\epsilon\times|\mu_{X}|^\rho\), $\sigma_{X}\sim \mathrm{Unif}(0,1)$, $\sigma_{Y}\sim\mathrm{Unif}(1,2)$ and  \(\epsilon \sim  \mathcal{N}(0,1)\).
\end{itemize}

\begin{figure}[t]
\centering
\includegraphics[width=\textwidth]{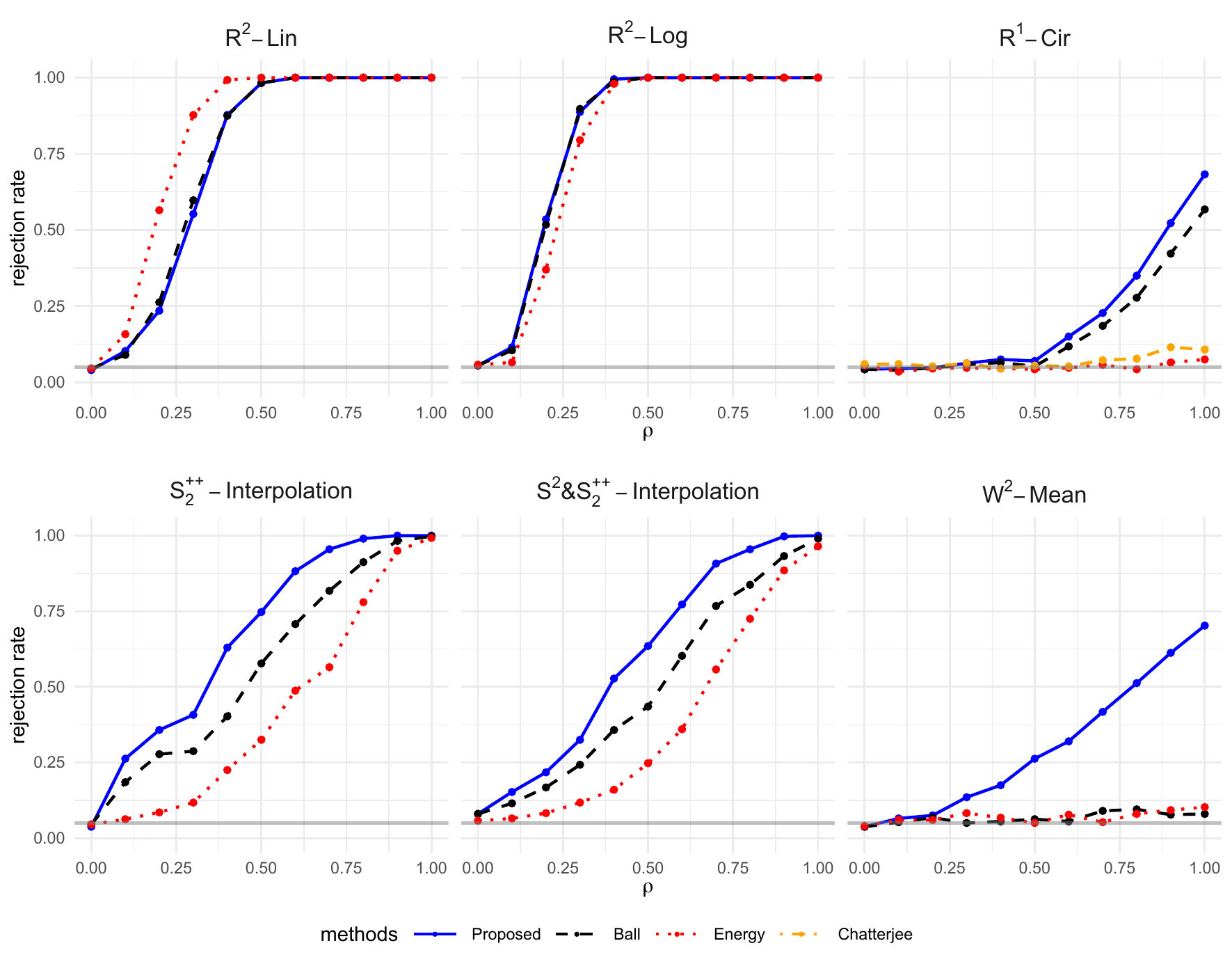}
\caption{Power curves comparing the proposed profile-based independence test (blue solid lines), the ball divergence  test (black dashed lines), the energy test (red dotted lines) and the Chatterjee test (yellow dash-dotted lines), where power is plotted as a function of the dependency parameter $\rho$, defined in the second paragraph of section \ref{s:simu}, across six simulation settings:
\textbf{$\mathbb{R}^{1}$-Cir}, \textbf{$\mathbb{R}^{2}$-Log}, and \textbf{$\mathbb{R}^{10}$-Log} (left, middle, and right panels in the first row);
\textbf{$\mathcal{S}_{2}^{++}$-Interpolation}, \textbf{$\mathbb{S}^2$\&$\mathcal{S}_{2}^{++}$-Hybrid}, \textbf{$\mathbb{W}^2$-Mean} (left, middle, and right panels in the second row).
The dashed grey line indicates the target significance level of 0.05, and the sample size is set to $n = 200$.}
\label{fig:main_f}
\end{figure}

We compare the proposed profile-based independence test with the energy test \citep{szekely2007measuring} and the ball divergence test \citep{pan2020ball}. All methods are implemented using available \texttt{R} packages: the energy test via \texttt{dcov.test()} from the \texttt{energy} package,  the ball divergence test via \texttt{bcov.test()} from the \texttt{Ball} package \citep{zhu2021ball}, and  Chatterjee's rank test via \texttt{xicor()} through the \texttt{XICOR} package.
Since Chatterjee's rank test only supports scalar independence testing, we include it only for the $\mathbb{R}^{1}$ setting.
Figure \ref{fig:main_f} presents power curves as functions of the dependency parameter $\rho$ for the simulation settings described above. 
The simulation results show that the proposed profile-based test effectively adapts to complex dependency structures where traditional methods tend to fail.
For instance, in the Euclidean space $\mathbb{R}^2$, when $Y$ and $X$ exhibit linear dependence, the energy test slightly outperforms the profile-based test. However, the proposed method performs better when the dependency becomes logarithmic.
In the more complex setting \textbf{\(\mathbb{R}^{1}\)-Cir}, both the energy test and Chatterjee test fail, while the proposed test achieves good power and outperforms the ball divergence test.
Moreover, in more challenging scenarios involving non-Euclidean spaces $\mathcal{S}_{2}^{++}$, $\mathbb{S}^2$, and $\mathbb{W}^2$, the proposed test performs significantly better than all other comparable methods.
These results are consistent with our theoretical findings and highlight the robustness and efficiency of the proposed test across a wide range of dependency structures. 
In summary, the proposed method performs comparably to existing approaches under linear dependence and outperforms them when the dependency structure is more intricate.

\begin{figure}[ht]
\centering
\includegraphics[width=\textwidth]{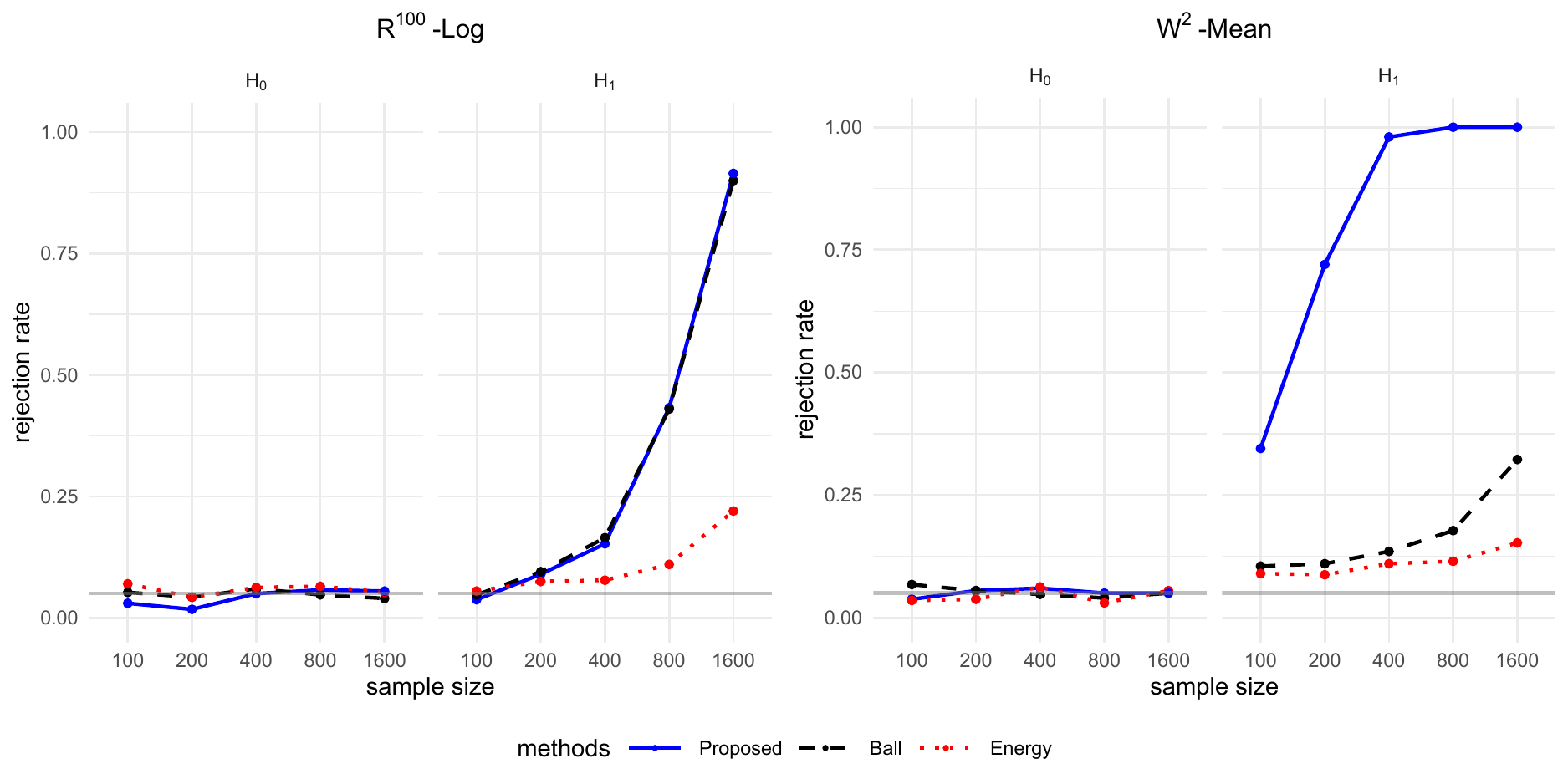}
\caption{Rejection rate of the proposed independence test for two high-dimensional settings (left two panels: \textbf{$\mathbb{R}^{100}$-Log}, right two panels: \textbf{$\mathbb{W}^{2}$-Mean}). Different colors and linetypes represent different methods: blue solid for the proposed method, black dashed for the ball divergence test, and red dotted for the energy test. Within each setting, two separate plots show the results under $\mathcal{H}_0$ (left) and $\mathcal{H}_1$ (right). The dashed grey line indicates the target significance level of 0.05 }
\label{fig:highD}
\end{figure}

We next study the performance of the proposed method in several high-dimensional settings. 
Figure \ref{fig:highD} presents the rejection rates as a function of sample size for the settings \textbf{$\mathbb{R}^{100}$-Log} and \textbf{$\mathbb{W}^{2}$-Mean}  under both $\mathcal{H}_0$ and $\mathcal{H}_1$. 
The results demonstrate that the proposed method consistently controls the type I error across all sample sizes.
In the Euclidean \textbf{$\mathbb{R}^{100}$-Log} setting, the proposed method outperforms the energy test and performs comparably to the ball divergence test, with slightly better power for large sample sizes. 
This suggests that, although the proposed method is not specifically tailored for high-dimensional data, it still achieves competitive or superior finite-sample performance.
In the infinite-dimensional \textbf{$\mathbb{W}^{2}$-Mean} setting, the proposed method exhibits a near-perfect power curve for relatively large sample sizes, while all other methods fail and show limited rejection rates.

We include results for additional settings in  the Supplement, illustrating how the power curves evolve with increasing sample size and approach perfect power for large samples. 
Additionally, we assess the performance of the proposed test across a range of significance levels in the Supplement.

\begin{figure}[ht]
\centering
\includegraphics[width=\textwidth]{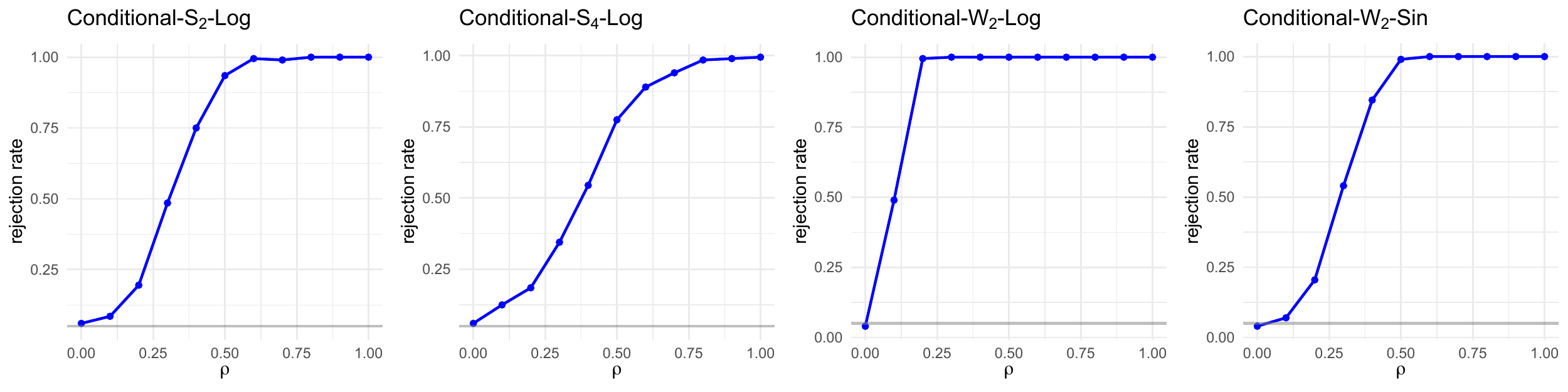}
\caption{Power curves of the proposed conditional independence test across four simulation settings with different metric spaces: \textbf{Conditional-$\mathbb{S}^{2}$-Log}, \textbf{Conditional-$\mathbb{S}^{4}$-Log}, \textbf{Conditional-$\mathbb{W}^{2}$-Log}, and \textbf{Conditional-$\mathbb{W}^{2}$-Sin}. Each plot shows the rejection rate as a function of the dependency parameter $\rho$, defined in the second paragraph of section \ref{s:simu}, with sample size fixed at $n=200$. The covariate $Z$ is drawn from $\mathrm{Unif}[0,1]$, and each point is based on 500 Monte Carlo replications. The dashed grey line indicates the target significance  level of 0.05. }
\label{fig:cond_main}
\end{figure}

{For the conditional independence testing, since there are no existing methods designed specifically for random objects, we focus on evaluating the performance of the proposed method in several  metric space settings. 
In all simulation settings below, the covariate $Z$ is generated from a $\mathrm{Unif}[0,1]$ distribution, and $\rho$ denotes the strength of dependence as defined in the second paragraph of section \ref{s:simu},}
\begin{itemize}
	\item \textbf{Conditional-$\mathbb{S}^{p}$-Log}: $\bm{X}=(X_1, X_2, \ldots, X_{p+1})$ and $\bm{Y}=(Y_1, Y_2, \ldots, Y_{p+1})$ are random elements on the unit sphere $\mathbb{S}^{p}$, generated with $Z$ and independent standard normal random variable $\epsilon_1,\,\epsilon_2,\ldots,\epsilon_{2p+1}$:
	$$X_{1}=\frac{Z+\epsilon_{1}}{\|\bm \epsilon_{X} \| },\, \,X_{2}=\frac{\epsilon_{2}}{\|\bm \epsilon_{X} \|},\,\ldots, \,X_{p+1}=\frac{\epsilon_{p+1}}{\|\bm \epsilon_{X} \|}; $$  
	$$Y_{1}=\frac{Z+\rho\log(4\epsilon_1^2)  + (1-\rho)\epsilon_{p+2} }{\|\bm \epsilon_{Y}\|}, \, \,Y_{2}=\frac{\epsilon_{p+3}}{\|\bm \epsilon_{Y} \|},\,\ldots,\, \,Y_{3}=\frac{\epsilon_{2p+2}}{\|\bm \epsilon_{Y} \|},$$  
	where $\bm\epsilon_{X}=(Z+\epsilon_{1},\epsilon_{2},\ldots,\epsilon_{p+1} )^{\top} $, $\bm\epsilon_{Y}=( Z+\rho\log(4\epsilon_1^2)  + (1-\rho)\epsilon_{p+2},\epsilon_{p+3},\ldots,\epsilon_{2p+2} )^{\top}$.  
	\item \textbf{Conditional-$\mathbb{W}^{2}$-Log}: \(X\) and \(Y\) are random elements in  the space of distributions \(\mathbb{W}^2\) with the Wasserstein metric, specifically  \(X =\mathcal{N}(\mu_{X},1)\) and \(Y=\mathcal{N}(\mu_{Y},1)\), where \(\mu_{X} \sim  Z+\mathrm{Unif}(-1,1)\),  \(\mu_{Y}=Z+\rho\log(4\mu_{X}^2) + \epsilon \), and  \(\epsilon \sim  \mathrm{Unif}(-1,1)\) and is independent of \(\mu_{X}\) and $Z$.
	\item \textbf{Conditional-$\mathbb{W}^{2}$-Sin}: \(X\) and \(Y\) are random elements in  the space of distributions \(\mathbb{W}^2\) with the Wasserstein metric, specifically  \(X =\mathcal{N}(\mu_{X},1)\) and \(Y=\mathcal{N}(\mu_{Y},1)\), where \(\mu_{X} \sim  Z+\mathrm{Unif}(-1,1)\),  \(\mu_{Y}=Z+\rho\sin(\mu_{X}\pi) + \epsilon \), and  \(\epsilon \sim  \mathrm{Unif}(-1,1)\) and is independent of \(\mu_{X}\) and $Z$.
\end{itemize}
{The results are shown in Figure \ref{fig:cond_main}. The proposed conditional test shows increasing power as the dependency parameter $\rho$ grows, achieving near-perfect rejection rates for moderate to strong dependence. 
These results highlight the adaptability and robustness of the proposed method across different metric spaces and across  complex nonlinear scenarios such as the log and sinusoidal dependence structures.}

\section{Data illustrations}\label{s:data}
\subsection{Connectivity networks in fMRI based brain imaging}\label{s:hcp}

Functional MRI (fMRI) imaging data acquisition is a well-established and important tool for the study of  brain function, due to its high temporal resolution. It is deployed to  capture  spontaneous neural activity during the  resting state and also for task-specific responses to external stimuli. 
In general, fMRI produces  complex data that require extensive pre-processing and then can be viewed as  connectivity matrices, random densities or signals that reflect  intricate spatial and temporal dynamics.
Here we  apply the proposed profile independence test  to investigate the relationship between resting-state and task-evoked fMRI brain connectivity using data from the HCP 1200 subject release \citep{van2013wu, elam2021human}.
Understanding this association reveals how the brain's functional architecture varies across different tasks compared to its intrinsic organization that is thought to be reflected during the resting state.

The data are  publicly accessible at \url{https://www.humanconnectome.org}. For our analysis, we focus on a subset comprising \( n = 209 \) young healthy adults who completed both resting-state and motor task fMRI sessions.
The resting-state fMRI data were collected while participants were instructed to keep their eyes open and fixate on a projected bright cross-hair on a dark background. 
For  the motor task fMRI recordings, participants were presented with visual cues prompting them to perform specific movements, such as squeezing their left or right toes or moving their tongue. 
Each time block for the recording of  a specific movement type lasted 12 seconds, included 10 movements and was preceded by a 3-second cue.
After preprocessing,  the average blood-oxygen-level-dependent (BOLD) signals were recorded across different voxels at 284 time points.
Further details about the experiment, preprocessing  and data acquisition are available in the reference manual for the WU-Minn HCP 1200 Subjects Data Release \citep{wu20171200}.

Our analysis focuses on nine regions of interest,  nucleus accumbens, amygdala, caudate, cerebellum, diencephalon, hippocampus, pallidum, putamen, and thalamus, in both left and right brain hemispheres. 
Correlation matrices of BOLD signals were constructed as per established practice in fMRI analysis, namely calculating the normalized cross-correlations of fMRI time series based on the voxels residing around a seed region for each region of interest, separately for left and right brain hemispheres. 
For the task-evoked fMRI signals,  the correlation matrices were calculated using the time segments corresponding to each specific task. For instance, the correlation matrix for ``squeezing left toe'' was derived from the time periods during which participants were actively performing the ``squeezing left toe'' task.
{For each subject, this resulted in a total of four \( 18 \times 18 \) correlation matrices for the whole brain, corresponding to one rest phase and three task phases (resting state, squeezing left toe, squeezing right toe, and moving tongue).}

\begin{figure}[t]
 	\centering
 	\includegraphics[width=0.99\textwidth]{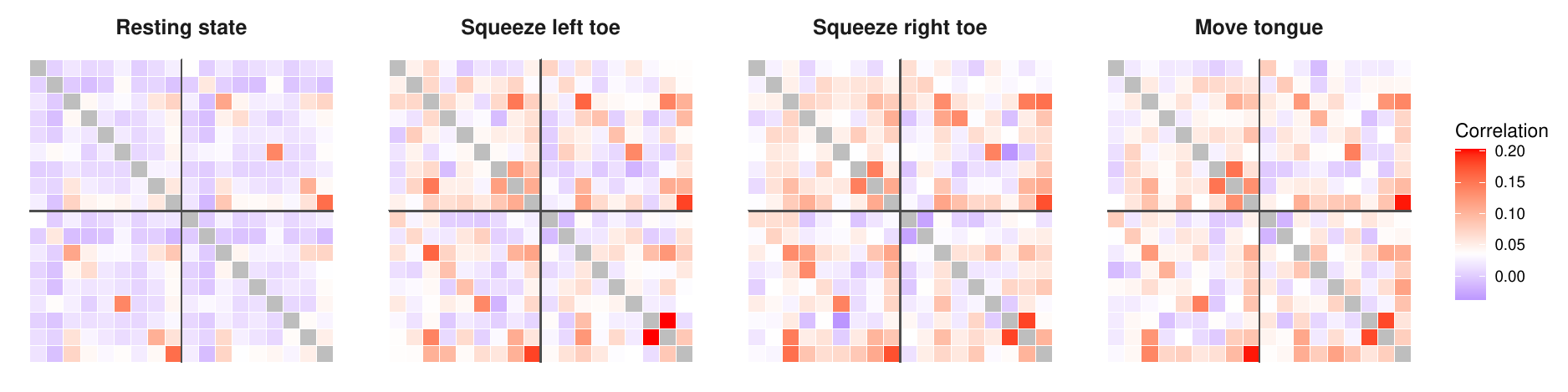}
 	\caption{Fr\'echet mean correlation matrices for resting-state fMRI (leftmost column) and task-evoked fMRI (next three columns: squeezing the left toe, squeezing the right toe, and moving the tongue). In each matrix, the top-left $9 \times 9$ block represents the connectivity within the left hemisphere, and the bottom-right $9 \times 9$ block represents the connectivity within the right hemisphere. }
 	\label{f:hm}
 \end{figure}

{
Figure \ref{f:hm} shows the estimated Fr\'echet mean of the four covariance matrices under the Frobenius norm. 
The results indicate that task-evoked correlation matrices exhibit stronger correlations across different brain regions compared to resting-state correlation matrices. 
Furthermore, in both resting-state and task-evoked conditions, the connectivity matrices between the left and right hemispheres are highly correlated. Applying both the proposed independence test and the energy test to these hemispheric connectivity matrices yields $p$-values below 0.01 in all cases, confirming that the left and right brain connectivities are not independent, which is consistent  with expectations.}

{
Given the strong  dependence between left and right hemisphere connectivity matrices observed across both resting-state and task-evoked conditions, we focus our  analysis on separate $9 \times 9$ intra-hemispheric correlation matrices. 
This is motivated by the fact that when left and right hemispheres are highly correlated, performing independence tests on the full $18 \times 18$ whole-brain matrices can make it easier to reject the null hypothesis of independence due to inflated signal from inter-hemispheric correlations. 
In contrast, tests based on the smaller $9 \times 9$ matrices are more stringent and informative for evaluating within-hemisphere dynamics. 
To this end, we applied the proposed  test and the energy test \citep{szekely2007measuring} to assess the independence of random object pairs $\{(X_{i}, Y_{i})\}_{i=1}^{209}$.
Here, $X_i$ denotes the correlation matrix from the resting-state fMRI, and $Y_i$ denotes the correlation matrix from the task-evoked fMRI  corresponding to the various motor tasks, for the left and right hemispheres of the $i$th subject.
We consider three metrics for the space of $9 \times 9$ positive definite matrices $A$ and $B$. Specifically, we compare the results for the following metrics that  were  applied previously in the literature for various data analysis tasks:  }
\begin{itemize}
    \item[a)] \textbf{Frobenius metric} 
    \[
    d_F({A}, {B}) = \|A - B\|_{F} = \sqrt{\operatorname{tr}\left[({A} - {B})({A} - {B})^\top\right]}.
    \]
    \item[b)] \textbf{Affine Invariant Riemannian Metric (AIRM)}  \citep{moakher2005differential}
    \[
    d_{R}(A, B) = \|\log(A^{-1}B)\|_{F}.
    \]
    \item[c)] \textbf{Log-Cholesky metric} \citep{lin2019riemannian}
    \[
    d_{C}(A, B) = \left\{\left\|\lfloor L_A \rfloor - \lfloor L_B \rfloor \right\|_{F}^2 + \sum_{j=1}^{9} (\log L_{A,jj} - \log L_{B,jj})^2 \right\}^{1/2},
    \]
    where $L_A$ and $L_B$ are the Cholesky decompositions of $A$ and $B$, respectively. Here $\lfloor L_A \rfloor$ denotes the off-diagonal part of $L_A$ and $L_{A,jj}$ the $j$th diagonal element of $L_A$.
    \item[c)] \textbf{$\alpha$-Power metric} \citep{dryden2009, zhou2022network}
    $$d_{F,\alpha}=\|A^{\alpha}-B^{\alpha}\|_{F},$$
    where $A^{\alpha}=U\Lambda^{\alpha}U^{\top}$ and $A=U\Lambda U^{\top}$ is the usual spectraL decomposition of a positive definite matrix $A$.
    \item[d)] \textbf{Bures-Wasserstein metric}
    $$
d_{{BW}}(A, B)^2 = \operatorname{Tr}(A + B - 2 (A^{1/2} B A^{1/2})^{1/2}).
$$
\end{itemize}

For each metric and method, we conducted six tests in total: three for each motor task (``squeezing the left toe'', ``squeezing the right toe'', and ``moving the tongue'') separately for the left and right hemispheres.
The $p$-values resulting from the permutation test implementation of the proposed profile independence test  are shown in Table \ref{t:hcp} and are also compared with the results from the energy test for independence. 
For the tongue movement task, both the  proposed profile independence test and the energy test yield the same result across all three metrics: there is no evidence to reject the null hypothesis that the correlation matrices between the resting state and the tongue movement task are independent.
For the toe-squeezing task, under the Frobenius metric, both the proposed profile independence test  and the energy test indicate dependence at the 0.05 significance level. However, the energy test fails to detect dependence under the AIRM, Log-Cholesky, $1/2$-Power, $1/3$-Power and Bures--Wasserstein metrics for the right toe-squeezing task, whereas the  proposed profile independence test consistently rejects the null hypothesis in all these cases.
These results demonstrate the consistency of the proposed test across different metric choices.

For completeness, we also include test results based on the full $18 \times 18$ correlation matrices in the Supplement, with the above caveat. Both the proposed method and the energy test yield consistently low $p$-values for the left and right toe-squeezing tasks, whereas the $p$-values for the tongue movement task are comparatively higher. These results are consistent across all metrics.

\begin{table}[tb]
    \centering
    \caption{Comparison of $p$-values obtained when applying the proposed profile independence test and the energy test for  independence for the null hypothesis of independence between resting-state fMRI and task-evoked fMRI connectivity matrices for three motor tasks: $\mathrm{L toe}$ (squeezing the left toe), $\mathrm{R toe}$ (squeezing the right toe), and $\mathrm{tongue}$ (moving the tongue). The results are presented for left and right hemispheres and for five different metrics:  Frobenius metric  (left three columns, upper table), Affine-invariant Riemannian metric  (middle three columns, upper table), Log-Cholesky metric  (right three columns, upper table), $1/2$-Power metric (left three columns, lower table), $1/3$-Power metric (middle three columns, lower table)  and Bures--Wasserstein metric (right three columns, lower table). The first two rows in upper/lower table show the results for the correlation matrices of the  left hemisphere and the last  two sub-rows for  the right hemisphere. The  $p$-values for the proposed profile independence test and the  energy test were obtained through permutation tests with a total number of permutations $N=500$ and are bolded if $p>0.1$. }
    \begin{tabular}{clrrrrrrrrrrrr}
        \toprule
        \multicolumn{1}{l}{ } & \multirow{2}[4]{*}{} & \multicolumn{3}{c}{Frobenius metric} & &\multicolumn{3}{c}{AIRM metric} && \multicolumn{3}{c}{Log-Cholesky metric} \\ 
        \cmidrule{3-13}
        \multicolumn{1}{l}{} &       & \multicolumn{1}{l}{Ltoe} & \multicolumn{1}{l}{Rtoe} & \multicolumn{1}{l}{tongue}& & \multicolumn{1}{l}{Ltoe} & \multicolumn{1}{l}{Rtoe} & \multicolumn{1}{l}{tongue}& & \multicolumn{1}{l}{Ltoe} & \multicolumn{1}{l}{Rtoe} & \multicolumn{1}{l}{tongue} \\ 
        \midrule
        \multirow[c]{2}{*}{Left} & proposed & 0.00  & 0.03  & {\bf 0.26}  & &0.00  & 0.01  & {\bf 0.78} & & 0.00  & 0.01  & {\bf 0.52} \\ 
        \cmidrule{2-13}
        & energy & 0.01  & 0.05  & {\bf 0.48} & & 0.01  & {\bf 0.48}  & {\bf 0.51}  & & 0.02  & {\bf 0.30}  & {\bf 0.54} \\ 
        \midrule
        \multirow[c]{2}{*}{Right} & proposed & 0.04  & 0.04  & {\bf 0.45} & & 0.02  & 0.03  & {\bf 0.75} & & 0.03  & 0.04  & {\bf 0.77} \\ 
        \cmidrule{2-13}
        & energy & 0.00  & 0.00  & {\bf 0.13} & & 0.00  & 0.00  & {\bf 0.20}  & & 0.02  & 0.02  & {\bf 0.22} \\ 
        \bottomrule
    \end{tabular}\\
    \begin{tabular}{clrrrrrrrrrrrr}
        \toprule
        \multicolumn{1}{l}{ } & \multirow{2}[4]{*}{} & \multicolumn{3}{c}{1/2-Power metric} & &\multicolumn{3}{c}{1/3-Power metric} && \multicolumn{3}{c}{Bures-Wasserstein} \\ 
        \cmidrule{3-13}
        \multicolumn{1}{l}{} &       & \multicolumn{1}{l}{Ltoe} & \multicolumn{1}{l}{Rtoe} & \multicolumn{1}{l}{tongue}& & \multicolumn{1}{l}{Ltoe} & \multicolumn{1}{l}{Rtoe} & \multicolumn{1}{l}{tongue}& & \multicolumn{1}{l}{Ltoe} & \multicolumn{1}{l}{Rtoe} & \multicolumn{1}{l}{tongue} \\ 
        \midrule
        \multirow[c]{2}{*}{Left} & proposed & 0.00  & 0.03  & {\bf 0.55}  & &0.00  & 0.02  & {\bf 0.68} & & 0.00  & 0.03  & {\bf 0.54} \\ 
        \cmidrule{2-13}
         & energy & 0.01  & {\bf 0.25}  & {\bf 0.55} & & 0.01  & {\bf 0.37}  & {\bf 0.57}  & & 0.01  & {\bf 0.26}  & {\bf 0.55} \\  
        \midrule
        \multirow[c]{2}{*}{Right} & proposed & 0.02  & 0.03  & {\bf 0.42} & & 0.01  & 0.02  & {\bf 0.49} & & 0.02  & 0.03  & {\bf 0.40} \\  
        \cmidrule{2-13}
        & energy & 0.00  & 0.00  & {\bf 0.22} & & 0.01  & 0.00  & {\bf 0.25}  & & 0.01  & 0.00  & {\bf 0.22} \\  
        \bottomrule
    \end{tabular}
    \label{t:hcp}%
\end{table}

\subsection{Age-at-death distributions}\label{s:hls}
The study of human mortality as quantified by age-at-death distributions  has attracted much interest over the past decades. The analysis of age-at-death distributions has led to deeper  insights into mortality patterns, lifespan variability and the factors influencing survival across different populations and time periods.
The Human Mortality Database (\url{www.mortality.org}) provides  annual age-at-death tables for 38 countries, with death rates recorded for each country across ages ranging from 0 to 110. 
By applying basic smoothing techniques to the yearly life tables available  in this database, one may obtain smooth age-at-death density distributions indexed by both country and calendar year.
We focus on 33 countries with available data spanning the years 1983 to 2018 and employ the conditional framework detailed in Section \ref{s:cond},  where the data are considered to be triplets \((X, Y, Z)\). Here  \(X\)  stands for the  male age-at-death distribution  for a given country at  calendar year \(Z\), and \(Y\) the  corresponding age-at-death distribution for females for the same country and year.  
{We use calendar time as covariate $Z$, an inherently continuous variable, although it is initially observed only at discrete time points, taken to be the midpoints of calendar years.  
With 35 consecutive years in the dataset and mortality rates exhibiting smooth temporal variation, this observation scheme  allows for the effective application of local smoothing methods in the pre-processing.}
We employ  the Wasserstein-2 metric \eqref{wass}  in the space of one-dimensional age-at-death distributions, as it is particularly sensitive to horizontal shifts in distributions that are of special interest. 

We applied  the  proposed conditional profile association \(\Delta_{XY}(z)\) in \eqref{d:cpda0} for this analysis, aiming  to shed light on the changing nature of the relationship between male and female age-at-death distributions  across different calendar years. The conditional profile association serves to   quantify the degree of dependence between the two distributions as a function of calendar time  \(z\), where a smaller value of \(\Delta_{XY}(z)\) indicates that male and female death rate distributions are relatively less dependent, and vice versa.  

\begin{figure}[t]
 	\centering
 	\includegraphics[width=0.9\textwidth]{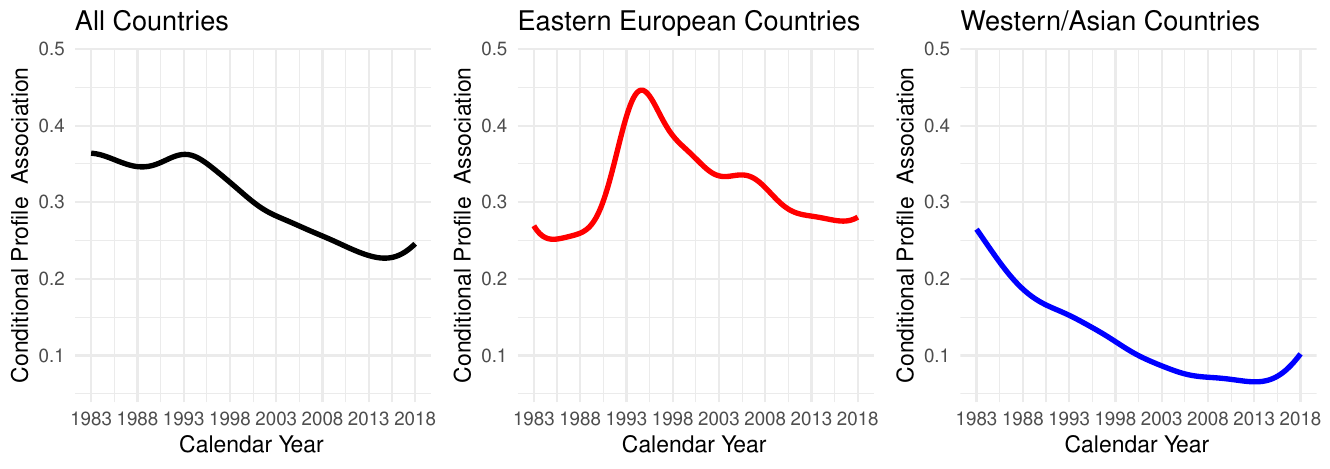}
 	\caption{Estimated conditional profile associations as a function of calendar year  for distributional data that correspond to male and female age-at-death distributions across 33 countries.   Conditional profile associations are shown for all  countries (left panel) and separately for Eastern European countries  (middle panel) and  Western/Asian countries (right panel). The associations are rescaled to lie between 0 and 1.}
 	\label{f:cpa}
 \end{figure}

We divided the 33 countries into two groups: Eastern European countries (Bulgaria, Belarus, Czechia, Estonia, Hungary, Lithuania, Latvia, Poland, Slovakia, and Slovenia)  and  Western/Asian countries (Australia, Austria, Belgium, Canada, Switzerland, Denmark, Spain, Finland, France, England and Wales, Greece, Ireland, Iceland, Italy, Japan, Luxembourg, the Netherlands, Norway, New Zealand, Portugal, Sweden, Taiwan and the United States). 
The results are in 
Figure \ref{f:cpa}, which illustrates  the estimated conditional profile association as a function of calendar year for all countries combined and  separately for Eastern European and Western/Asian countries. 

The left panel, representing all countries, exhibits a generally decreasing trend over time, indicating a gradual reduction in the association between male and female age-at-death distributions over the 25 years covered in this analysis. 
The middle panel, which focuses on Eastern European countries only, shows a noticeable peak in the early 1990s, suggesting a period of heightened dependence between male and female mortality distributions, potentially due to major socio-political and economic changes in the region during that time that had a strong effect on both female and male age-at-death distributions. After this peak, the association declines but remains relatively stable. 
In contrast, the right panel, which represents Western/Asian countries, shows a consistently lower conditional profile association compared to both all countries combined and Eastern European countries. 
The association in Western countries also follows a steady downward trend, suggesting an increasing independence between male and female mortality distributions over time. 
This lower and declining association seems to indicate an increasing decoupling between male and female mortality over time that is more expressed for Western/Asian countries. 
The clear disparity between Western and Eastern European countries highlights the varying  demographic, political  and health-related forces  that influence the relationship between  male and female mortality.

{To complement the observed decreasing trend in conditional profile associations between male and female mortality, we conducted the proposed conditional independence test described in Section \ref{s:cond}. 
The resulting $p$-values are consistently below 0.01 for both Eastern European and Western/Asian countries, providing strong statistical evidence against conditional independence over the time period considered. 
These findings are expected and confirm that mortality rates for males and females are overall dependent, as expected.}

\section{Discussion}\label{s:dicu}

In this paper, we introduce the profile  association, a  measure designed to quantify associations between random objects. 
The proposed association is based on distance profiles, which serve as bridges connecting  the distributions of random objects, the classical Hoeffding's D statistic, and the  proposed association measure. 
Utilizing this framework, we develop an independence testing procedure for random objects, demonstrating superior performance compared to existing methods in numerical studies.
We further extend the profile  association and the associated testing framework to conditional settings, in which random objects are paired with Euclidean predictors. 
This extension provides a useful  tool for exploring relationships between complex object responses and Euclidean covariates, with immediate practical relevance. 
Beyond independence testing, the profile  association framework has  potential for broader statistical applications, including causal inference. 

Causal inference has emerged as an important  field in statistics, with significant applications across various scientific disciplines. 
One key challenge in causal inference is detecting confounding contributions that can lead to  spurious associations arising from the influence of a confounder on 
two outcomes of interest. 
{A simple conceptual tool for understanding confounding for Euclidean data is the classical law of total covariance, 
$$
\text{Cov}(X, Y) = \mathbb{E}[\text{Cov}(X, Y \mid Z)] + \text{Cov}(\mathbb{E}[X \mid Z], \mathbb{E}[Y \mid Z]),
$$
which decomposes the conditional covariance between random variables \( X \) and \( Y \) in the presence of a conditioning variable $Z$. Here 
 \( \text{Cov}(\mathbb{E}[X \mid Z], \mathbb{E}[Y \mid Z]) \) is the contribution to the covariance explained by $Z$, and thus possibly due to a confounding effect of $Z$,  whereas  $\mathbb{E}[\text{Cov}(X, Y \mid Z)]$ 
 is the part that is not explained by $Z$ in the context of a linear framework. }
A question for future research  is whether by utilizing the proposed conditional profile association one can extend such a separation of the effect that can be attributed to $Z$ and the remaining association that cannot be explained by $Z$ to  scenarios where \( X \) and \( Y \) are random objects and a linear framework does not apply, due to the lack of vector operations in non-Euclidean spaces. 

Considering  the mortality example discussed in Section \ref{s:hls}, where  $X_{i}$ and  $Y_{i}$ are female and male age-at-death distributions, while  $Z_{i}$ is a Euclidean covariate, a preliminary attempt to shed some light on this is the following three-step procedure:   first compute the unconditional profile association $\Delta(X,Y)$,  using all available data pairs $\{(X_{i},Y_{i})\}_{i}$; second, replace the conditional expectations $\mathbb{E}[X \mid Z]$,  $\mathbb{E}[Y \mid Z]$ in the law of total covariance  by  Fr\'echet regressions \citep{petersen2019frechet} 
to obtain conditional \F means  $\mathbb{E}_{\oplus}[X\mid Z]$ and $\mathbb{E}_{\oplus}[Y\mid Z]$ using data  $\{(X_{i},Z_{i})\}_{i}$ and $\{(Y_{i},Z_{i})\}_{i}$ respectively; third, 
compute  $\Delta(\mathbb{E}_{\oplus}[X\mid Z], \mathbb{E}_{\oplus} [Y\mid Z])$ by plugging in the results from the \F regression step. 

Applying these considerations for the human mortality data for all countries combined  yields $30\Delta(X,Y)=0.36$ for the estimate of the overall profile association that ignores calendar year, but  $30\Delta(\mathbb{E}_{\oplus}[X\mid Z], \mathbb{E}_{\oplus} [Y\mid Z])=0.97$ for the estimate of the part of the association explained by $Z$.  Here we introduce the multiplier $30$ to normalize the association values so they lie in  the interval $[0,1]$. Thus when calendar year is ignored, the observed profile association between male and female age-at-death distributions is much weaker than the value obtained when one first uses the calendar year as predictor. 
This discussion is not meant to be conclusive but rather to showcase the potential for future statistical research on theory and applications of conditional profile association for random objects.

\section*{Acknowledgments}
This research was supported in part by NSF DMS-2310450. We thank Professor Russell Lyons, Indiana University, for important comments.

\bigskip
\bibliographystyle{Chicago}
\bibliography{report.bib}

%
%


\end{document}